\documentclass[epj,nopacs]{svjour}
\usepackage{bm}        
\usepackage{amssymb}  
\usepackage{dsfont}
\usepackage{xcolor}
\usepackage{nccfoots}
\usepackage{epsfig}
\usepackage{epstopdf}
\definecolor{eq}{rgb}{190,190,190}
\usepackage{cuted}
\usepackage{graphics}
\usepackage{graphicx}  
\usepackage{dcolumn}   
\usepackage{amsmath}
\usepackage[latin9]{inputenc}
\usepackage[flushleft]{threeparttable}
\usepackage{multicol}
\usepackage{threeparttable}
\usepackage{float}
\usepackage{dsfont}
\usepackage{array}
\usepackage{multirow}
\usepackage[square,numbers,sort&compress]{natbib}
\usepackage{caption}
\usepackage{mathrsfs} 
\usepackage{textgreek}

\begin{document}

\title{Concepts for direct frequency-comb spectroscopy of $^{229\text{m}}$Th\\ and an internal-conversion-based solid-state nuclear clock}
\titlerunning{Concepts for direct frequency-comb spectroscopy of $^{229\text{m}}$Th}
												
\author{Lars von der Wense\inst{1,2} \and Chuankun Zhang\inst{2}}

\institute{Ludwig-Maximilians-Universit\"at M\"unchen, 85748 Garching, Germany \and JILA and Department of Physics, University of Colorado, Boulder, CO 80309-0440, USA }

\date{\today}


\abstract{
A new concept for narrow-band direct nuclear laser spectroscopy of $^{229\text{m}}$Th is proposed,
using a single comb mode of a vacuum ultraviolet frequency comb generated from the 7th harmonic of an Yb-doped fiber laser system. In this concept more than $10^{13}$ $^{229}$Th atoms on a surface are irradiated in parallel and a successful nuclear excitation is probed via the internal-conversion (IC) decay channel. A net scanning time of 15 minutes for the most recent 1~$\sigma$ energy uncertainty interval of 0.34~eV appears to be achievable when searching for the nuclear transition. In case of successful observation, the isomer's energy value would be constrained to an uncertainty of about 100~MHz, which is a factor of $10^6$ of improvement compared to today's knowledge. Further, the comb mode could be stabilized to the nuclear transition using the same detection method, allowing for the development of an IC-based solid-state nuclear clock, which is shown to achieve the same performance as a crystal-lattice nuclear clock, however, with the advantage of a drastically simpler detection scheme. Finally, it is shown that the same laser system could be used to narrow down the isomer's transition energy by further six orders of magnitude during laser excitation of $^{229}$Th$^{3+}$ ions in a Paul trap and to drive nuclear Rabi oscillations, as required for the development of a nuclear clock based on a single $^{229}$Th$^{3+}$ ion.
}

\maketitle

\section{\label{sec1}Introduction}
\noindent The accuracy of time measurements has considerably evolved within the past decades, starting with the development of the first atomic clock by Louis Essen and Jack Parry in 1955 \cite{Essen}, followed by the legal definition of the second by means of the Cesium standard in 1967 \cite{Cesium} and the invention of the fountain clock in the late 1980s \cite{Kasevich}. With the invention of the optical frequency comb \cite{Reichert,Udem,Stephen} a new technological leap has occurred, allowing for the development of optical atomic clocks \cite{Ludlow}, which by today pose the most precise timekeepers. The achieved accuracies of these clocks vary around $10^{-18}$, corresponding to an error of 1 second in 30 billion years, significantly longer than the age of the universe \cite{Nicholson,Hunteman,McGrew,Brewer}. Such accuracy in frequency measurement opens up a plethora of applications, e.g., in satellite-based navigation \cite{Gill}, geodesy \cite{Mehlstaeubler} and the search for temporal variation of fundamental constants \cite{Rosenband,Godun}.\\[0.2cm]
As an alternative to the existing optical atomic clocks, the idea of a nuclear optical clock was developed \cite{Peik,Campbell,Rellergert,Kazakov}. Compared to an optical atomic clock, the operational principle remains unchanged, however, with the important difference that a nuclear transition instead of an atomic shell transition is used for time measurement. A nuclear optical clock is expected to outperform even today's best atomic clocks due to conceptual advantages. An accuracy approaching $10^{-19}$ and therefore about an order of magnitude of improvement compared to the best optical atomic clocks operational today was estimated for a nuclear clock based on a single $^{229}$Th ion \cite{Campbell}. In a pictorial understanding, the reason for this expected improved accuracy is that the atomic nucleus is several orders of magnitude smaller than the atomic shell and therefore, due to the related smallness of the nuclear moments, drastically more stable against external influences \cite{Peik2}.\\[0.2cm]
Historically, the idea of a nuclear clock dates back to the late 1950s, shortly after the discovery of the Mössbauer effect \cite{Moessbauer}. At that time it was noticed by Pound and Rebka that the concept of Mössbauer spectroscopy, due to its potential to measure extremely small relative energy differences, could be used to observe the gravitational red-shift \cite{Pound}. Here, a 93.3~keV nuclear excited state of $^{67}$Zn, which exhibits an extraordinary narrow relative linewidth of $\Delta f/f=5.5\cdot10^{-16}$, is of particular interest. Mössbauer spectroscopy with $^{67}$Zn was first achieved in 1960 together with the proposal for a $^{67}$Zn-based nuclear clock \cite{Nagle,Katila}. However, in this context, the expression ``nuclear clock" refers to a relative frequency measurement rather than to an absolute frequency determination.\\[0.2cm]
A nuclear optical clock has not yet been experimentally realized. One challenge is that most nuclear transitions exhibit an excitation energy which is several orders of magnitude above what is accessible with today's narrow-bandwidth laser technology. However, there is one exception, which is the first nuclear excited state of $^{229}$Th, known as the ``thorium isomer" ($^{229\text{m}}$Th). With an excitation energy that has recently been measured to be about 8.3~eV \cite{Seiferle2019}, $^{229\text{m}}$Th is the nuclear excited state of lowest known excitation energy. In fact, the energy is within reach of currently available frequency comb technology thereby in principle allowing for nuclear laser spectroscopy and the development of a nuclear optical clock. The state is metastable, with an estimated radiative lifetime of about $10^4$ seconds \cite{Tkalya2015,Minkov2019}. If non-radiative decay channels are suppressed, e.g., in individual $^{229}$Th$^{3+}$ ions, the long radiative lifetime would lead to a narrow relative linewidth of up to $\Delta f/f\approx10^{-20}$, which allows the development of an optical atomic clock of highest stability \cite{Peik,Campbell}. In the following, the experimental history of $^{229\text{m}}$Th will briefly be sketched.\\[0.2cm]
$^{229}$Th was first considered to possess a low-energy nuclear excited state in 1976, when it helped explaining some peculiarities of the $^{229}$Th nuclear $\gamma$-ray spectrum as observed in the $\alpha$ decay of $^{233}$U, which remained otherwise unexplained \cite{Kroger}. Based on $\gamma$-ray spectroscopy of nuclear rotational states of larger energies, the $^{229\text{m}}$Th energy value was constrained to $-1\pm4$~eV in 1990 \cite{Reich1} and then further improved to $3.5\pm1$~eV in 1994 \cite{Helmer}, which was until 2007 the adopted energy value. In 2007, however, a different energy value was measured using a metallic magnetic microcalorimeter with an improved energy resolution leading to $7.6\pm0.5$~eV \cite{Beck1}, later corrected to $7.8\pm0.5$~eV \cite{Beck2}.\\[0.2cm]
A central problem, that has hindered the development of a nuclear clock based on $^{229\text{m}}$Th within the past decade is, that its energy is not sufficiently well constrained to allow for laser spectroscopy of individual laser-cooled $^{229}$Th ions. This has led to a multitude of efforts to pin down the isomeric energy value to higher precision (a recent review can be found in Ref.~\cite{Wense2018}). It would be ideal to measure the energy via spectroscopy of photons directly emitted in the isomer's ground state decay. However, despite worldwide efforts, until today no unambiguous signal of photons emitted in the isomeric decay has been observed, potentially pointing toward a significant non-radiative decay channel \cite{Jeet,Yamaguchi,Wense3,Stellmer}.\\[0.2cm]
In 2016, the direct observation of electrons emitted in the internal conversion (IC) decay channel of the isomeric state opened a new path for exploration of the isomer's properties \cite{Wense1}. In the IC decay, the nucleus couples to the atomic shell and transfers its energy to an electron, which is subsequently ejected. Importantly, the nucleus couples directly to the atomic shell and the process does not correspond to a $\gamma$ decay followed by photo-electron emission. For this reason the lifetime of the nuclear excited state shortens by the amount of the conversion coefficient $\alpha_\text{ic}$, if IC is energetically permitted. Based on the observation of the IC electrons, the $^{229\text{m}}$Th IC-lifetime in neutral, surface-bound atoms was determined to be about 10~$\mu$s \cite{Seiferle2}, in agreement with theoretical expectations \cite{Strizhov,Karpeshin1,Tkalya2015}. Further, in 2018, collinear laser-spectroscopy of the electronic shell of $^{229}$Th$^{2+}$ ions with 2\% in the isomeric state led to the observation of the isomer-induced change of the hyperfine-structure \cite{Thielking}. Independently, also a new way of isomer population via excitation of the 29~keV state in $^{229}$Th has been experimentally realized \cite{Masuda}. This allowed for a $^{229\text{m}}$Th energy determination to $8.30\pm0.92$~eV in \cite{Yamaguchi2019}. In 2019, the $^{229\text{m}}$Th energy was also measured to higher precision via direct spectroscopy of the IC electrons emitted in the isomeric decay, thereby drawing a clear path for future nuclear laser spectroscopy experiments \cite{Seiferle2019}. The value obtained by this method is $8.28\pm0.17$~eV ($149.7^{+3.2}_{-3.0}$~nm) and is used throughout this paper. More recently the Beck et al micro-calorimetric measurement was repeated with improved detector resolution, resulting in a value of $8.09^{+0.14}_{-0.19}$~eV ($153.3^{+3.6}_{-2.7}$~nm) \cite{Geist2020}. This value would even be advantageous for the proposed experiment as it is closer to the center wavelength of the here considered 7th harmonic of an Yb-doped fiber frequency comb (152.8~nm).\\[0.2cm]

\section{Concepts for nuclear laser excitation}
Various concepts for nuclear laser excitation of $^{229\text{m}}$Th were discussed in literature. Two different principles require particular attention: (1) The isomer's excitation via the electronic bridge (EB) excitation mechanism and (2) direct laser excitation of the nucleus. Both approaches will be discussed individually in the following.

\subsection{The electronic bridge excitation mechanism}
In the EB excitation process, the hyperfine coupling between the atomic shell and the nucleus is used for nuclear excitation (see, e.g., Ref. \cite{Campbell2011}). This allows the excitation of a (virtual) electronic shell state instead of directly exciting the nucleus. During the decay of the excited electronic shell state, part of its energy is transferred to the nucleus with a certain probability, resulting in nuclear excitation. Historically, the process was introduced by Morita and Batkin \cite{Morita,Batkin}. The reverse process, which is the de-excitation of a nuclear state via the excitation of an electronic shell state, was already discussed earlier by Krutov \cite{Krutov}. The process was applied to the excitation of $^{229\text{m}}$Th by E.V. Tkalya in 1992 \cite{Tkalya1992}. Later, the concept was considered for various different physical conditions (see, e.g., Refs.~\cite{Kalman1994,Karpeshin1996,Tkalya1996,Matinyan,Karpeshin1999,Porsev2010,Campbelldiss,Karpeshin2015,Karpeshin2017,Bilous2018,Andreev2019,Mueller2019}). Importantly, any direct nuclear laser excitation will most likely be accompanied by a process of nuclear excitation via EB. In general, the EB process is expected to have larger excitation rates compared to direct laser excitation, as the electronic shell acts as a resonator, thereby enhancing the electric field at the site of the nucleus. Enhancement factors between $10$ and $10^4$ were predicted in Ref. \cite{Porsev2010}, depending on the isomer's exact energy value and a potential overlap with an electronic shell transition. Electronic bridge excitation has not yet been experimentally observed, however the reverse process, EB decay, was observed for $^{93}$Nb \cite{Kekez} as well as $^{193}$Ir \cite{Zheltonozhskii} and the closely related process of nuclear excitation by electron transition (NEET) has been detected for $^{197}$Au \cite{Kishimoto2000} and $^{193}$Ir \cite{Kishimoto2005}.\\[0.2cm] 
In 2012 it was proposed to excite $^{229\text{m}}$Th with a two-photon EB excitation scheme using a frequency comb \cite{Romanenko}. This concept is intriguing, as in this way all comb modes would contribute to the nuclear excitation. Further, a laser system operating at the technologically easier accessible wavelength of about 300~nm could be used. The proposed excitation scheme makes important use of the EB mechanism, as the probability for direct nuclear two-photon excitation can generally be considered as very low \cite{Bilous2018}. The investigation of the EB effect and the potential for an EB-based nuclear clock, that may even include nuclear transitions other than $^{229\text{m}}$Th \cite{Berengut2018}, is an interesting field of continued research.\\[0.2cm]
In this paper, purely direct nuclear excitation is considered, leading to a conservative estimate for the nuclear excitation rates. This is mainly owed to the complexity of the EB process, which will drastically depend on the electronic shell and the details of the excitation scheme.\\[0.2cm]

\subsection{Direct nuclear laser spectroscopy}
Direct nuclear laser spectroscopy of $^{229}$Th was first considered in the 1990s \cite{Tkalya1992,Tkalya1996,Karpeshin1999}, followed by the proposal for a nuclear clock in 2003 \cite{Peik}. At that time, it was found to be unfavorable compared to other nuclear laser excitation schemes that make use of the EB mechanism. Direct laser excitation of $^{229\text{m}}$Th is challenging for three reasons: (1) The natural linewidth corresponding to the radiative decay channel is extraordinary narrow, $\Gamma_\gamma=10^{-4}$~Hz when assuming a radiative lifetime $\tau_\gamma$ of $10^4$~s. Therefore, the fraction of laser light that contributes to the nuclear excitation can generally considered to be small. (2) The isomer's excitation energy is only weakly constrained, which leads to a large energy range that has to be scanned when searching for the nuclear excitation. (3) The energy is located in the vacuum-ultra-violet (VUV) energy range, where laser technology is not easily available.\\[0.2cm]
Despite these challenges, direct laser spectroscopy of $^{229\text{m}}$Th has become a viable option within the recent years, due to technological advances in laser technology (see, e.g., Ref.~\cite{Zhang2020} and references therein), an improved knowledge about the isomer's transition energy \cite{Seiferle2019} and new concepts for nuclear laser excitation \cite{Wense2}.\\[0.2cm]
Due to compactness, robustness and stability, it would be most advantageous to use a continuous-wave (cw) laser for $^{229\text{m}}$Th nuclear excitation. The $^{229\text{m}}$Th energy value of $8.28\pm0.17$~eV, corresponding to a wavelength of $149.7\pm3.1$~nm, makes this, however, challenging. Unfortunately, no cw laser at about 150~nm wavelength is available by today. Due to lack of a broadband gain medium in the VUV region, building a laser oscillator in the VUV is not possible right now. Therefore, nonlinear conversion like second harmonic generation (SHG) would be required for obtaining a cw laser in the VUV. This can only be achieved with significant intensity in non-linear crystals. The only available non-linear crystal that allows for tunable cw-light generation in the deep VUV is KBe$_2$BO$_3$F$_2$ (KBBF) \cite{Chen}. Although transparent down to 147~nm \cite{Chen} and already being used for the generation of pulsed laser light at below 150~nm via sum frequency mixing \cite{Nakazato}, KBBF fulfills the phase-matching condition required for SHG only down to 164~nm \cite{Halasyamani}. Alternative crystals for cw-light generation at 150~nm might be BaMgF$_4$ (BMF) \cite{Shimamura} and BPO$_4$ \cite{Zhang2011}. Both crystals provide good transparency at 150~nm, however, they do not offer the possibility for phase matching. Quasi-phase-matching (QPM) might, however, be possible \cite{Villora}. Recently, improved computational methods also allowed the prediction of a couple of new materials that may offer the possibility for SHG at 150~nm, these are AlCO$_3$F, PB$_3$O$_6$F$_2$ and SiCO$_3$F$_2$ \cite{Kang}. For these reasons there is hope, that cw laser technology for nuclear laser excitation of $^{229\text{m}}$Th will be available in the future. However, as long as this is not the case, other methods have to be investigated.\\[0.2cm]
Direct frequency-comb spectroscopy \cite{Cingoez,Ozawa2017} could be a promising alternative way for narrow-band nuclear laser excitation. During direct frequency-comb spectroscopy, a single comb line is in resonance with the nuclear transition and the energy is determined from knowledge of the repetition frequency $f_\text{rep}$ and the carrier envelope offset frequency $f_\text{ceo}$ via the comb equation $f_\text{nucl}=f_\text{ceo}+N_\text{nucl}f_\text{rep}$ \cite{Picque}. The comb mode number $N_\text{nucl}$ can be obtained from multiple measurements with different repetition rates \cite{Zhang2007}. In recent years, the frequency-comb technology has been drastically extended to higher energies and now also covers wavelengths in the deep to extreme ultra-violet region (see, e.g., Refs.~\cite{Gohle2005,Jones2005,Ozawa2008,Yost2009,Cingoez,Kobayashi,Ozawa2013,Pupeza,Pupeza2014,Carstens,Benko,Ozawa2015,Saule2018,Porat,Saule2019,Mills2019,Seres2019,Zhang2020}). The method that is usually applied for VUV frequency-comb generation is cavity-enhanced high-harmonic generation (HHG) in a noble gas. In particular, the 7th harmonic of an Yb-doped fiber laser \cite{Cingoez,Pupeza,Ozawa2017,Zhang2020} or the 5th harmonic of a Ti:Sapphire laser \cite{Seres2019} are matching with the $^{229\text{m}}$Th nuclear transition energy.\\[0.2cm]
In this paper, the potential to perform direct frequency-comb spectroscopy of $^{229\text{m}}$Th is discussed. It is assumed that the the 7th harmonic of an Yb-doped fiber laser, generated via cavity-enhanced harmonic generation, is used for nuclear excitation. Parameters of an existing laser system operational at the JILA laboratory in Boulder, Colorado \cite{Cingoez,Benko,Porat,Zhang2020} are used  for the quantitative estimates. Two different approaches to use this laser for nuclear spectroscopy are discussed: In the first concept the irradiation of more than $10^{13}$ $^{229}$Th atoms on a surface is analyzed, which can be considered as the simplest method and would allow for a high-precision $^{229\text{m}}$Th energy determination and the development of an IC-based solid-state nuclear clock. In the second concept, nuclear frequency-comb spectroscopy of laser-cooled $^{229}$Th$^{3+}$ ions in a Paul trap is considered. It is found, that the laser intensity would be sufficient to drive nuclear Rabi oscillations in case that the bandwidth of an individual comb mode is narrowed down to the Hz-range as would be ideal for clock operation.


\section{Theoretical Background}
\noindent In the following, the nuclear excitation probability as a function of time under resonant laser irradiation will be modeled.
For this purpose, a closed nuclear two-level system consisting of ground-level sub-state $g$ and excited level sub-state $e$ is assumed.
In case of laser excitation of a large number of $^{229}$Th atoms on a surface, as discussed in Sec.~\ref{sec2}, a closed nuclear two-level system is a valid assumption, as the nuclear de-excitation rate via the IC-channel is large, not leading to any significant depopulation of the ground-level. In fact, the only reason why the system cannot be modeled via the Einstein rate equations, as was done in Ref.~\cite{Wense2}, is, that the bandwidth of the laser light used for excitation is not significantly broader than the natural linewidth of the IC-broadened nuclear transition. For the case of laser excitation of $^{229}$Th ions in a Paul trap, as considered in Sec.~\ref{concept2sec}, the laser-induced depopulation of the excited state is significantly larger than the natural decay rate, again validating the two-level approximation.\\[0.2cm]
Under this assumption, the time-dependent nuclear excitation probability $\rho_\text{exc}(t)$ for a single nucleus under resonant irradiation is given by Torrey's solution of the optical Bloch equations \cite{Wense2020,Noh}:\\

\begin{strip}
\begin{equation}
\label{Torrey}
\rho_\text{exc}(t)=\frac{\Omega_{eg}^2}{2\left(\Gamma\tilde{\Gamma}+\Omega_{eg}^2\right)}\left[1-e^{-\frac{1}{2}(\Gamma+\tilde{\Gamma})t}\left(\cos(\lambda t)+\frac{\Gamma+\tilde{\Gamma}}{2\lambda}\sin(\lambda t)\right)\right].
\end{equation}
\end{strip}

\noindent Here $t$ denotes the time since the start of the irradiation, $\Omega_{eg}$ is the Rabi frequency specific to the considered sub-states, $\Gamma=\Gamma_\gamma+\Gamma_\text{nr}$ is the total decay rate of the nuclear excited state, including $\gamma$ decay of rate $\Gamma_\gamma$ as well as potential non-radiative decay channels $\Gamma_\text{nr}$. In case of a non-vanishing IC decay channel, which is expressed by a non-zero IC coefficient $\alpha_\text{ic}=\Gamma_\text{ic}/\Gamma_\gamma$, one obtains $\Gamma_\text{nr}=\Gamma_\text{ic}$ and therefore $\Gamma=(1+\alpha_\text{ic})\Gamma_\gamma$. The parameter $\lambda$ is defined as $\lambda=\lvert\Omega_{eg}^2-(\tilde{\Gamma}-\Gamma)^2/4\rvert^{1/2}$. For $\Omega_{eg}<\lvert\tilde{\Gamma}-\Gamma\rvert/2$, the $\sin$ and $\cos$ functions in Eq.~(\ref{Torrey}) have to be exchanged by $\sinh$ and $\cosh$, respectively. Further,
\begin{equation}
\label{Gammatilde}
\tilde{\Gamma}=\frac{\Gamma+\Gamma_L}{2}+\tilde{\Gamma}_\text{add}
\end{equation}
 denotes the total decay rate of the coherences, with $\Gamma_L$ the bandwidth of the laser light used for excitation and $\tilde{\Gamma}_\text{add}$ corresponding to potential additional decoherence. Additional decoherence can arise due to homogeneous broadening in a solid-state environment as will be discussed in Sec.~\ref{Splitting} or due to, e.g., phonon coupling of ions in a Coulomb crystal. For laser-cooled $^{229}$Th$^{3+}$ ions, $\tilde{\Gamma}_\text{add}$ will in general be small and for the considered experimental conditions we assume $\tilde{\Gamma}_\text{add}\ll\Gamma_L$, which allows one to neglect additional decoherence. For a solid-state environment a value for $\tilde{\Gamma}_\text{add}$ between $2\pi\cdot 1$~kHz and $2\pi\cdot10$~kHz due to nuclear spin coupling was estimated in Ref.~\cite{Rellergert}, about $2\pi\cdot150$~Hz was obtained in Ref.~\cite{Kazakov}. Here it is assumed that additional decoherence in the solid-state environment is significantly smaller than the IC-enhanced natural $^{229\text{m}}$Th decay rate of $\Gamma\approx10^5$~Hz. This allows one to omit $\tilde{\Gamma}_\text{add}$ in all discussed calculations. Experimentally, this assumption is supported by the fact, that the $93.3$~keV Mössbauer state of $^{67}$Zn with 9~$\mu$s half-life could be resolved down to its natural linewidth of $2\pi\cdot12.4$~kHz \cite{Potzel}.\\[0.2cm]
The Rabi frequency of the system can be calculated as \cite{Wense2020}
\begin{equation}
\label{Rabi}
\Omega_{eg}=\sqrt{\frac{2\pi c^2 I C_{ge}^2\Gamma_\gamma}{\hbar\omega_0^3}},
\end{equation}
with $I$ the intensity of the laser light, $\omega_0$ the angular frequency of the nuclear transition, $\Gamma_\gamma$ the radiative decay rate of the nuclear excited level, $c$ the speed of light and $\hbar$ the reduced Planck constant.\\[0.2cm]
The factor $C_{ge}$ is a constant that depends on the specific sub-states of the nuclear ground- and excited level. In case of nuclei embedded in a solid-state environment, $C_{ge}$ is simply the Clebsch-Gordan coefficient corresponding to the transition. For nuclei of isolated atoms or ions the constant has a more complicated form that is given in Sec.~\ref{concept2sec}. As we do not want to consider any particular sub-state right in the beginning and are only interested in an order-of-magnitude estimate for the number of excited nuclei, we set $C_{ge}=1$ for the following calculations. It will be detailed in Sec. \ref{Splitting} that the actual Clebsch-Gordan coefficients for $^{229}$Th are $\sqrt{4/15}$ or $\sqrt{2/5}$. This will either lead to a reduction of the actually observed excitation rate by up to a factor of $4$ (for the case of $C_{ge}=\sqrt{4/15}$) or, in case of degeneracy of all sub-states, to an increase by a factor of $1.3$.\\[0.2cm]
Note, that Eq.~(\ref{Torrey}) remains valid also if the bandwidth of the laser light used for irradiation, $\Gamma_L$, is significantly broader than the total linewidth $\Gamma$ of the nuclear transition, as the definition of the decay rate of the coherences $\tilde{\Gamma}$ takes respect for that. In case of the low-saturation limit, which is fulfilled for $\tilde{\Gamma}\gg\Gamma\gg\Omega_{eg}$, Eq.~(\ref{Torrey}) transforms to
\begin{equation}
\rho_\text{exc}(t)=\frac{\Omega_{eg}^2}{2\tilde{\Gamma}\Gamma}\left(1-e^{-\Gamma t}\right).
\end{equation}
This result is consistent with the one obtained from the Einstein rate equations used in Ref.~\cite{Wense2}. Under this approximation, the nuclear excitation rate $\Gamma_\text{exc}$ can be estimated as $\dot{\rho}_\text{exc}(0)$ to be\footnote{Considering the relation between the absorption cross section $\sigma$ and the excitation rate, $\sigma=\hbar\omega_0\Gamma_\text{exc}/I$ \cite{Steck}, for $\Gamma_L\gg\Gamma$ one has $\sigma=\lambda^2/(2\pi)C_{ge}^2\Gamma_\gamma/\Gamma_L$, with $\lambda$ being the wavelength corresponding to the nuclear transition. This is in complete agreement with the expressions given in Refs.~\cite{Tkalya1992,Tkalya1996,Karpeshin1999}.}
\begin{equation}
\label{rate}
\Gamma_\text{exc}\approx \frac{\Omega_{eg}^2}{2\tilde{\Gamma}}=\frac{\pi c^2IC_{ge}^2\Gamma_\gamma}{\hbar\omega_0^3\tilde{\Gamma}}.
\end{equation}
The number of excited nuclei $N_\text{exc}$ is obtained from Eq.~(\ref{Torrey}) by multiplication with the number of irradiated nuclei $N_0$. Throughout the paper Eq.~(\ref{Torrey}) will be used to model the time-dependent nuclear excitation probability.

\section{\label{sec2} Laser excitation of $^{229\text{m}}$Th atoms on a surface}
\noindent In this section, it is proposed to use the 7th harmonic of an Yb-doped fiber frequency comb, to irradiate a solid sample of $^{229}$Th atoms deposited as a thin layer on a surface. The presented concept is based on a proposal presented in Refs.~\cite{Wense2,Wense4}, where it was shown that broadband VUV laser light, generated via four-wave mixing in a noble gas, could be used for $^{229\text{m}}$Th laser excitation if the isomer's IC decay channel is used to probe the nuclear resonance. Here, the concept is adapted to a source of narrow-band laser light as offered by a VUV frequency comb. The advantages of the presented scheme compared to the scheme proposed in Refs.~\cite{Wense2,Wense4} are that the isomeric energy could be constrained to higher precision and that the same concept could be used to develop a new type of nuclear-clock, which is based on the isomer's IC decay channel. Further, the same laser system could also be used for the excitation of $^{229}$Th$^{3+}$ ions in a Paul trap and for the development of a single-ion nuclear clock.

\subsection{General concept}
\begin{figure}[t]
\begin{center}
\includegraphics[width=8cm]{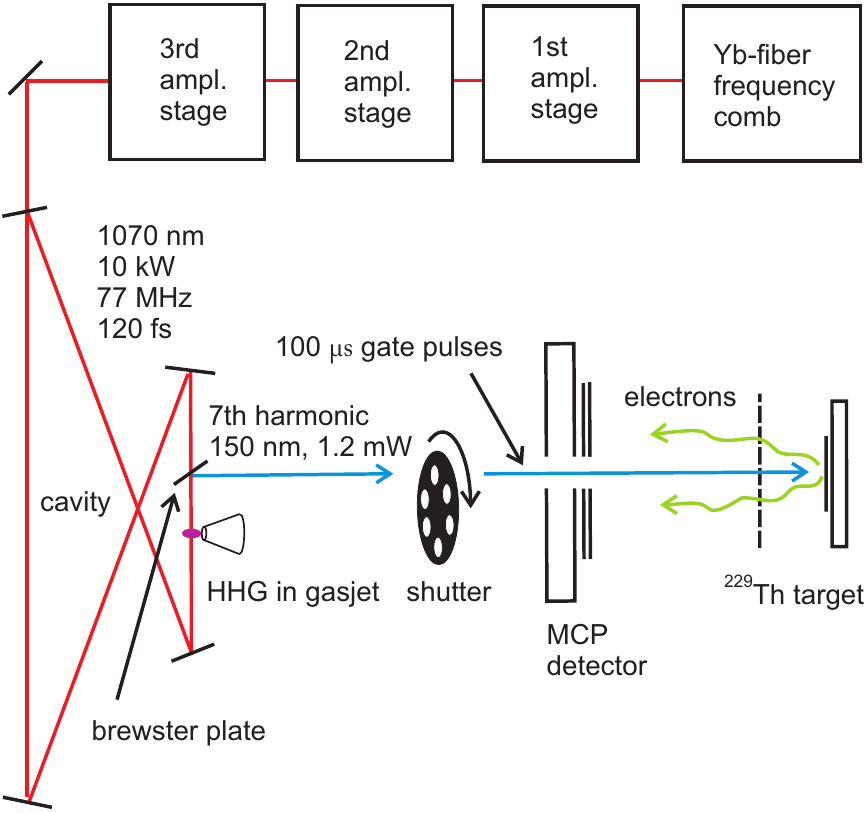}
\caption{\small Schematic of the experimental setup proposed for frequency-comb-based direct nuclear laser excitation of $^{229\text{m}}$Th. A vacuum ultra-violet (VUV) frequency comb is generated via cavity-enhanced harmonic generation from an Yb-doped-fiber frequency comb \cite{Cingoez,Benko,Porat,Zhang2020}. Gate pulses of 100 $\mu$s duration with 5 kHz gate frequency are produced with the help of a mechanical shutter \cite{Gembicky}. A solid sample surface of $^{229}$Th is irradiated with the laser light. In case of resonance, the nuclear isomeric state is excited and will decay within about 10~$\mu$s lifetime via internal conversion (IC) under emission of an electron \cite{Seiferle2}. The electrons can be guided with the help of electric fields and detected by an MCP detector.}
\label{fig1}
\end{center}
\end{figure}

The VUV frequency comb light consists of femtosecond pulses whose repetition rate determines the comb mode spacing. For the purpose of laser excitation of the $^{229}$Th nuclei, it is assumed that the laser light is gated, thereby generating fs pulse trains of 100~$\mu$s duration. In case that an individual mode of the frequency comb is tuned to the nuclear resonance energy, a fraction of the nuclei is excited into the isomeric state during a single gate pulse. After the end of the laser gate pulse the excited nuclei will decay with about 10~$\mu$s lifetime via internal conversion (IC) \cite{Wense1,Seiferle2}. The low-energy electrons emitted in the IC decay will be guided by electric fields and detected with the help of a micro-channel plate (MCP) detector \cite{Wiza} in order to probe the nuclear excitation. In this way it can be inferred if the laser light was tuned to the nuclear resonance. The absolute number of the comb mode that is in resonance with the nuclear transition and therefore also the absolute frequency can be obtained by making several measurements with different repetition rates (mode spacings) of the frequency comb \cite{Cingoez,Zhang2007}. A schematic of the experimental setup is shown in Fig.~\ref{fig1}. Important aspects of the scheme are: a huge number of irradiated atoms of more than $10^{13}$; an IC-broadened nuclear transition linewidth of 15.9~kHz and a 9 orders of magnitude shortened lifetime of about 10~$\mu$s, which allows for laser-triggered isomeric decay detection.
\noindent The concept requires three main components as visualized in Fig.~\ref{fig1}: A target consisting of a thin surface of $^{229}$Th, a gated laser system used for target irradiation and a detection system that will allow for the detection of the low-energy IC electrons with high efficiency. Each part will be discussed individually in the following.\\[0.2cm]

\subsection{The $^{229}$Th target} 
It is assumed, that the $^{229}$Th target consists of a $\sim10$~nm thin $^{229}$ThO$_2$ layer deposited as a round surface with 0.3~mm diameter onto a metal substrate. Technically it might be advantageous to start with a larger surface area and later cover the outer part of the target with a thin metal plate. ThO$_2$ is chosen as a target material, as it is chemically very stable and possesses a cubic lattice structure, which results in a vanishing electric quadrupole splitting of the nuclear states (see Sec.~\ref{Splitting}). The thorium oxide band gap was measured to about 5.8~eV \cite{Rodine}, which is significantly below the isomer's energy, therefore decay of the isomeric state will securely occur by internal conversion. Thin $^{229}$ThO$_2$ layers could be produced, e.g., by spray pyrolysis \cite{Mahmoud}, by photochemical deposition \cite{Huentupil,Arancibia}, by vapor deposition \cite{Bagge}, by the sol-gel technique \cite{Tripathi}, by the drop-on-demand technique \cite{Haas}, by hydrothermal synthesis \cite{Wang,Moeini} or from non-aqueous solution \cite{Hudry}. The useful thickness is limited by the mean-free-path length of electrons in the target material \cite{Seah} and the large absorption coefficient for VUV light \cite{Gillis}. With a ThO$_2$ density of 9.86~g/cm$^3$, the number of $^{229}$Th atoms in the target amounts to $1.6\cdot10^{13}$. As the $^{229}$Th half-life is about 7900 years, this corresponds to an activity of $\sim45$~Bq. Even if the target material was chemically purified prior to its deposition, within about 100 days, the $^{229}$Th decay chain approaches a radioactive equilibrium, leading to an activity increase by a factor of $8$. The target activity in equilibrium will therefore amount to about 360~Bq. This assumes that no other thorium isotopes are contained in the target material. In case of a potential contamination with $^{228}$Th, the $^{228}$Th activity may even exceed the one of $^{229}$Th due to the comparatively short $^{228}$Th half-life of 1.9 years. A $^{228}$Th contamination may easily occur, as $^{229}$Th is usually obtained by chemical separation from $^{233}$U material, which contains spurious amounts of $^{232}$U. Even in case that the fractional content of $^{228}$Th in the target material amounts to $10^{-3}$ only, its activity would be by a factor of 4.2 larger compared to that one of $^{229}$Th. Further, $^{228}$Th approaches the radioactive equilibrium with its daughters after about 20 days. The factor of activity enhancement is $7$. In this case, the total target activity in equilibrium would amount to about 1.7~kBq. Fortunately, the $^{228}$Th activity would fade away a few years after chemical separation from its mother nuclei.\\[0.2cm]
\begin{figure}[t]
\begin{center}
\includegraphics[totalheight=7cm]{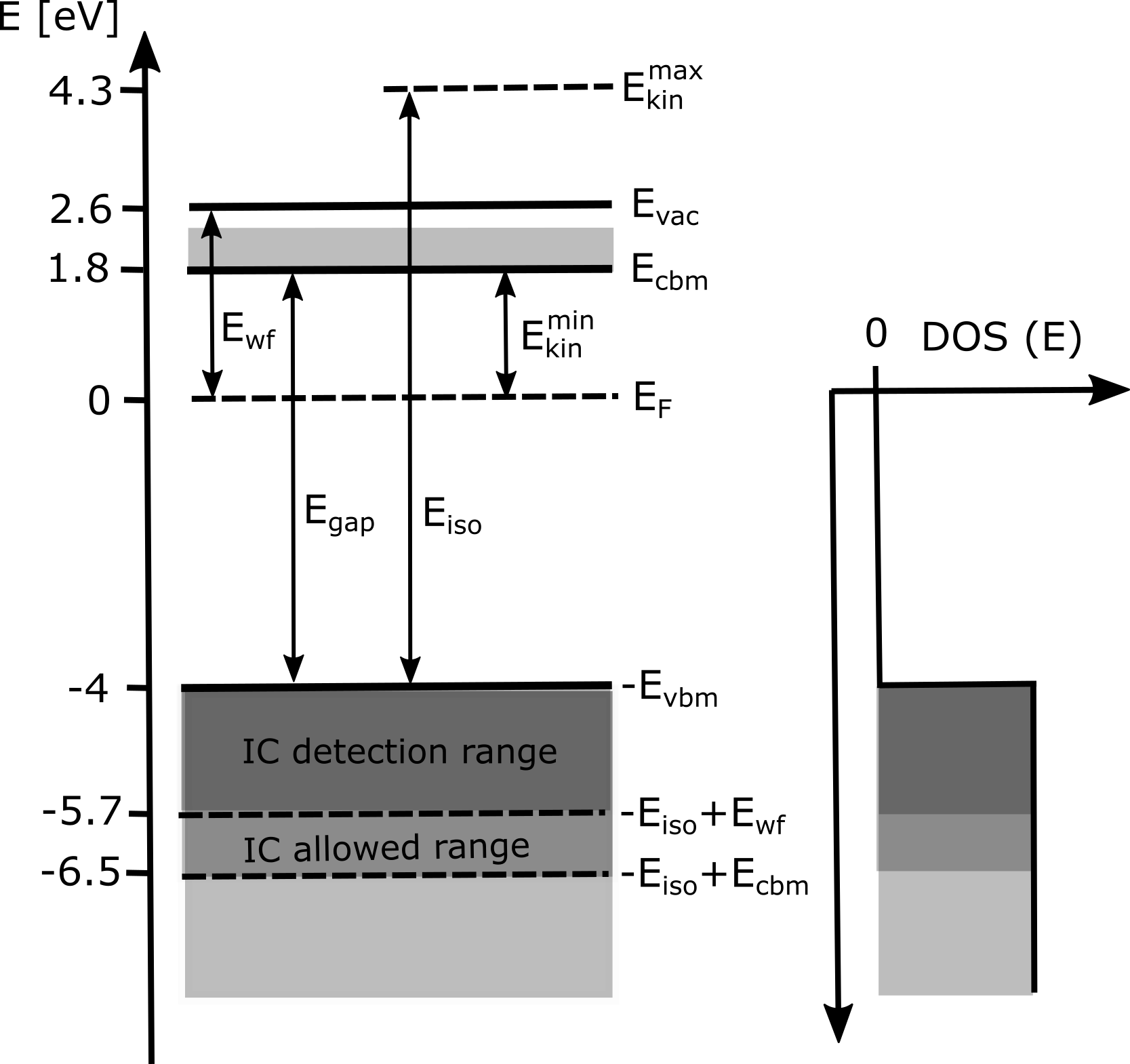}
\caption{\small Sketch of the ThO$_2$ band structure together with parameters used for the escape efficiency estimate: $E_F$ Fermi energy, $E_\text{vac}$ vacuum energy, $E_\text{wf}$ work function, $E_\text{cbm}$ conduction-band minimum, $E_\text{vbm}$ valence-band maximum, $E_\text{gap}$ band gap, $E_\text{iso}$ isomeric energy.}
\label{fermi}
\end{center}
\end{figure}
An important aspect for the proposed detection method is how efficient the low-energy IC electrons can leave the ThO$_2$ target. An analysis scheme for the escape efficiency from a metal surface was presented in Ref.~\cite{Verlinde}. Here we adapt the procedure to an insulator and take additionally also the quantum mechanical reflection of the electrons at the material surface into account. Input parameters are the inelastic mean-free-path (IMFP) length of the low-energy IC electrons in the target material, as well as the density-of-state (DOS) distribution of the electrons. For electrons being emitted at a certain depth of the target $h$, the escape efficiency can be estimated as \cite{Verlinde}
\begin{equation}
\label{escape}
\eta_\text{esc}(h)=\int_{-E_\text{iso}+E_\text{wf}}^{-E_\text{vbm}}\frac{\eta_\text{trans}(E)\eta_\text{surf}(E,h)\text{DOS}(E)dE}{\text{norm}}.
\end{equation}
Here $E_\text{iso}$ denotes the isomer's energy, $E_\text{wf}$ the ThO$_2$ workfunction and $E_\text{vbm}$ the energy of the valence-band maximum. All parameters are taken with respect to the Fermi level $E_F$ and are shown in Fig.~\ref{fermi}. $\eta_\text{trans}(E)$ denotes the quantum mechanical probability for the electron to be transmitted at the surface potential well. $\eta_\text{surf}(E,h)$ is the probability for an electron to reach the surface when it is emitted from an energy $E$ below the Fermi level at depth $h$ in the material and $\text{DOS}(E)$ is the density of states functional. ``$\text{norm}$" denotes a normalization factor defined as
\begin{equation}
\text{norm}=\int_{-E_\text{iso}+E_\text{cbm}}^{-E_\text{vbm}} \text{DOS}(E)dE,
\end{equation}
with $E_\text{cbm}$ being the energy of the conduction-band minimum. The lower limit of the integral takes respect for the fact that there are no electronic levels in the band gap that can be populated during the IC decay. Except for the additional transmission efficiency, Eq.~(\ref{escape}) transforms to the expression given in Ref.~\cite{Verlinde} for $E_\text{vbm}=E_\text{cbm}=0$ as would be the case in a metal. The transmission efficiency can be calculated as \cite{Eisberg}
\begin{equation}
\eta_\text{trans}(E)=\frac{4\sqrt{E_\text{kin}}\sqrt{E_\text{kin}-E_\text{wf}}}{\left(\sqrt{E_\text{kin}}+\sqrt{E_\text{kin}-E_\text{wf}}\right)^2}.
\end{equation}
Here the electron's kinetic energy with respect to the Fermi level $E_\text{kin}=E+E_\text{iso}$ was introduced. Further, the probability for the electron to reach the surface is obtained as \cite{Verlinde}
\begin{equation}
\eta_\text{surf}(E,h)=\frac{1}{\pi}\int_0^{\pi/2}\exp\left(\frac{-h}{\cos(\theta)\lambda_n(E_\text{kin})}\right)\sin(\theta)d\theta,
\end{equation}
where $\lambda_n(E_\text{kin})$ denotes the IMFP of the electrons in nanometer as a function of their kinetic energy above the Fermi level. Based on Ref.~\cite{Seah}, $\lambda_n$ is given as $\lambda_n=a\cdot\lambda_m$, with $a$ being the monolayer thickness of the material and $\lambda_m$ the IMFP in dimension of monolayers, which can be estimated for low-energy electrons propagating in an anorganic-compound material as $\lambda_m=2170/E_\text{kin}^2$ \cite{Seah}, where $E_\text{kin}$ denotes the electron energy above the Fermi level in eV. For ThO$_2$, the monolayer thickness is obtained as $a=\sqrt[3]{A/(\rho n N)}=0.24$~nm, where $A$ denotes the molecular mass, $\rho$ the density, $n$ the number of atoms in the molecule and $N$ the Avogadro constant. For electrons with an energy of 4.3~eV above the Fermi level, this leads to and IMFP of $28$~nm. Importantly, the IMFP is rising for electrons of lower energy, however, as soon as the energy is below the material's work-function, the electrons will not be able to leave the surface. For this reason, a high electron escape efficiency is obtained for materials with a low work function and a DOS distribution that exhibits a high density of electronic states around an energy of $E_\text{iso}-E_\text{wf}$ below the Fermi level, which corresponds to the lowest energy that still allows the electrons to overcome the work function. Fortunately, ThO$_2$ offers both, a very low work-function of only about 2.6~eV \cite{Bagge} and a high DOS between 4 and 6~eV below the Fermi level \cite{Cakir}.\\[0.2cm]
A further important parameter is the attenuation length of the VUV light in the ThO$_2$ target material. Experimental studies of the VUV absorption coefficient $\alpha$ of ThO$_2$ are scarce. A value of $\alpha\approx 0.1$~nm$^{-1}$ for light with 8.3~eV energy was found in Ref.~\cite{Gillis}. This translates into a transmission efficiency of 37\% for 10~nm layer thickness. The corresponding factor of decrease in nuclear excitation rate over the total depth of the target is $\eta_\text{abs}=0.63$. For this reason, significantly thicker ThO$_2$ layers are also not advantageous due to absorption losses in the material.\\[0.2cm]
The fraction of electrons that is able to leave the target material $\eta_\text{eff}$ can be numerically estimated based on the integral
\begin{equation}
\eta_\text{eff}=\frac{1}{d\eta_\text{abs}}\int_0^{d}\eta_\text{esc}(h)\text{e}^{-\alpha h} dh,
\end{equation}
with $d$ denoting the ThO$_2$ oxide layer thickness. It is found that the result does not significantly depend on the details of the DOS function, therefore a simple step function with a constant DOS starting at $E_\text{vbm}$ was used in the calculation. A relatively large escape efficiency of about 15\% is obtained for $d=10$~nm layer thickness.\\[0.2cm]
Using an insulator as a target material requires the discussion of charge-up effects during laser irradiation. Laser gate pulses of 100~$\mu$s duration with a total power of 1.2~mW ($1.2\cdot 10^5$ comb modes with 10~nW average power per mode) are assumed to be used for target irradiation, which corresponds to $9\cdot10^{10}$~photons at the considered wavelength of 150~nm. Under the pessimistic assumption that one electron is carried away by each photon, the total ThO$_2$ charge-up would amount to $Q=1.4\cdot10^{-8}$~C. Treating the ThO$_2$ layer deposited on a metal surface as a plate capacitor, the capacitance amounts to $C=2.2$~nF, where the dielectric constant $\epsilon_r=3.17$ of ThO$_2$ was used \cite{Mahmoud}. Therefore the voltage across the layer would be $U=6.4$~V. Now using the ThO$_2$ resistivity, which was measured to be about $10^6$~{\textOmega}cm \cite{Mahmoud}, the resistance of our thin layer amounts to about $R=130$~\textOmega. This in turn allows estimating the current across the layer to $\sim0.05$~A. The total charge would be compensated in about 0.3~$\mu$s, which is significantly shorter than the duration of a laser gate pulse. Therefore no charge accumulation has to be expected to occur during laser irradiation of the thin ThO$_2$ layer, which is in agreement with experimental observation \cite{Cakir}. 
 
\subsection{The laser system}
\begin{figure*}[t]
\begin{center}
\includegraphics[totalheight=5cm]{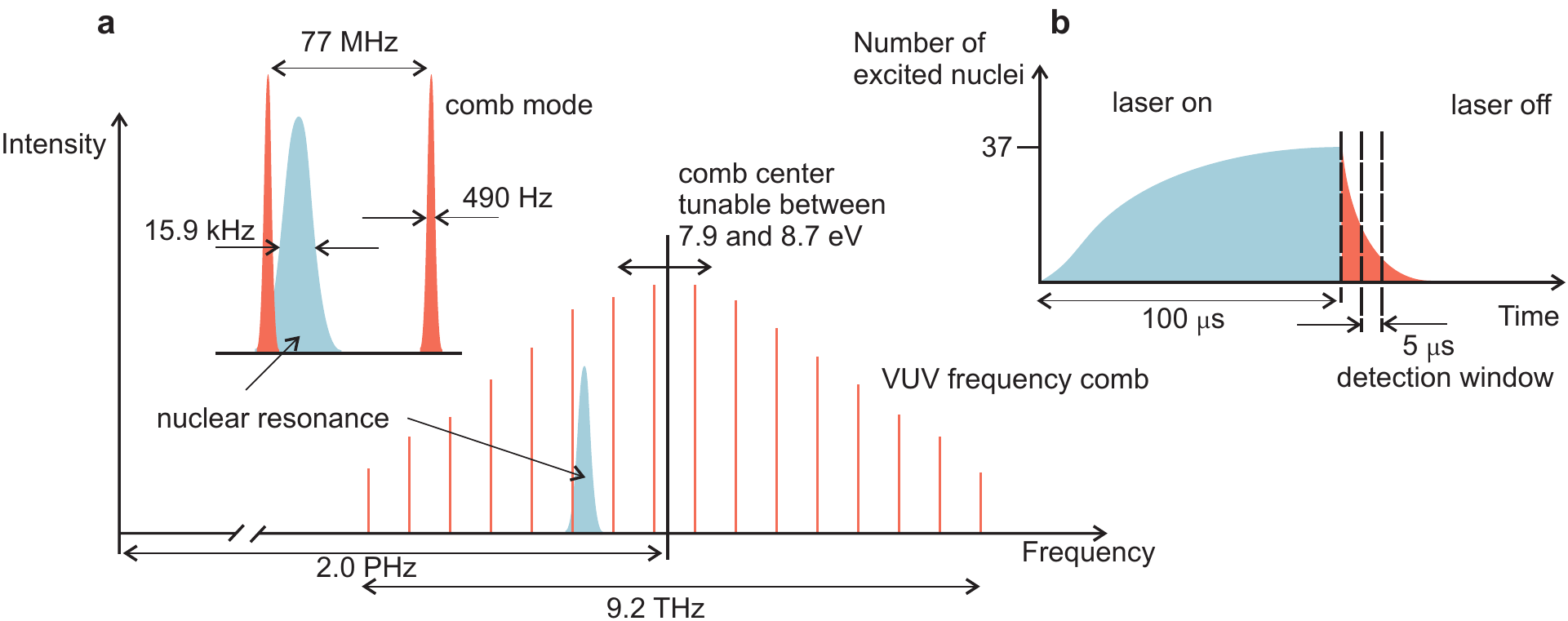}
\caption{\small Sketch of the proposed experimental concept together with important parameters. a) A VUV frequency comb with $1.2\cdot10^5$ comb modes of 490~Hz mode-width (10~nW power per mode) and 77~MHz mode spacing, tunable between 7.9 and 8.7~eV, is used to search for the nuclear resonance. b) Laser excitation is achieved during gate pulses of 100~$\mu$s duration (5 kHz gate frequency). The isomeric decay is observed shortly after the end of each laser pulse. A 5~$\mu$s detection window is assumed for a signal-to-background estimate.}
\label{fig2}
\end{center}
\end{figure*}

The laser light that is assumed to be used for irradiation of the $^{229}$Th target comes from the 7th harmonic of an ytterbium-doped fiber-based frequency comb operating at a wavelength around 1070~nm. Parameters of an existing frequency comb operational in the group of Jun Ye at JILA in Boulder, CO \cite{Cingoez,Benko,Porat,Zhang2020} are used for all quantitative estimates. Comparable laser systems are also available in Tokyo, Japan \cite{Ozawa2013,Ozawa2015,Kobayashi} and at MPQ in Germany \cite{Pupeza,Pupeza2014,Carstens,Saule2018,Saule2019}. With the help of an enhancement cavity, a high average power of 10~kW is produced. Focusing this laser power to a helium-xenon gas-jet leads to an intensity of $5\cdot10^{13}$~W/cm$^2$, sufficiently large to efficiently generate high harmonics. For the considered experiment, the 7th harmonic with a wavelength around 150~nm is of interest, which can be out-coupled with a high efficiency of up to 75\% for example with a Brewster plate \cite{Ozawa2015} or with a non-collinear out-coupling scheme \cite{Fomichev,Moll,Wu,Ozawa2008,Zhang2020}.
The seventh harmonic spans an energy range of 0.038~eV (9.2~THz) and tunability between 7.9 and 8.7~eV (143 to 157~nm) is planned to be provided. It consists of about $1.2\cdot10^5$ comb modes, each of them with a bandwidth of about 490~Hz and a power of about 10~nW. The mode spacing is 77~MHz. In the considered concept the laser beam is gated either with the help of a fast rotating shutter system \cite{Gembicky} or with an acousto-optical modulator to provide laser gate pulses of 100~$\mu$s pulse duration at 5~kHz gate frequency. Important parameters of the concept are visualized in Fig.~\ref{fig2}. The values of variables used for the quantitative analysis according to Eq.~(\ref{Torrey}) are listed in Tab.~\ref{table_surface}.\\[0.2cm]
When using this laser system for surface irradiation, the potential for laser ablation and surface heating requires consideration. The frequency comb has a repetition rate of 77~MHz and a fourier-transform limited pulse duration of about 50~fs of the 7th harmonic. Considering that all $1.2\cdot10^5$ comb modes will irradiate the target in parallel with a well defined phase relation to create a pulse train, the peak fluence of the light amounts to $\Phi=2.0\cdot10^{-8}$~J/cm$^2$. For estimating the target heating, two different time scales have to be considered. On a femtosecond timescale, no significant heat transport to the non-irradiated target substrate will take place and the absorbed energy leads to a local temperature rise of the target. Assuming that the light energy is completely transferred to heat, the maximum temperature rise per fs laser pulse is estimated as $\Delta T\approx \alpha\Phi/C_p=8.6\cdot10^{-3}$~K. Here $\alpha\approx 1\cdot10^{6}$~cm$^{-1}$ denotes the absorption coefficient of ThO$_2$ at 150~nm \cite{Gillis} and $C_p=2.33$~Jcm$^{-3}$K$^{-1}$ is the volumetric heat capacity of ThO$_2$ \cite{Osborne}. A heat-up of a few mK is negligible in terms of a potential isomer shift of the nuclear transition energy (see Sec.~\ref{Splitting}). The heat will diffuse fast (on the picosecond timescale \cite{Zhang}) and lead to an equilibrium temperature rise of the entire target substrate. Assuming that the target substrate consists of a Ti-sputtered Si-wafer of 0.5~mm thickness and 20~mm diameter and is thermally isolated from its environment, the temperature rise would be dominated by the specific heat capacity of Si of $C_p=1.66$~Jcm$^{-3}$K$^{-1}$, leading to $\Delta T=4.6$~mK per second of laser irradiation. A thermal contact with the environment is therefore required in order to keep the temperature rise significantly below 1~K during laser irradiation. Importantly, the laser fluence is eight orders of magnitude below the typical laser ablation threshold for femtosecond laser pulses \cite{Gamaly,Brown}, which is in agreement with the calculated temperature rise. For this reason no significant laser ablation of the target is expected to occur. However, carbon layer growth has to be considered and will require ultra-high vacuum conditions in the target chamber and potentially ozone purge (see Ref.~\cite{Wense2}).

\begin{table*}[t]
\begin{center}
\caption{Values of variables used for the calculation of the number of excited nuclei per laser gate pulse based on Eq.~(\ref{Torrey}).}
\begin{tabular}{clcl}
\hline\noalign{\smallskip}
Variable & Description & Value & Comment \\
\noalign{\smallskip}\hline\noalign{\smallskip}
$I$ & Laser intensity & $1.4\cdot10^{-5}$ W/cm$^{2}$ & Single-mode (10 nW) focused to $\varnothing$ $0.3$~mm\\
$\Gamma_L$ & Laser bandwidth & $2\pi\cdot490$ Hz & Smaller than IC-broadened nuclear linewidth\\
$\omega_0$ & Angular frequency & $2\pi\cdot2.0$ PHz & Corresponding to $8.3$~eV energy \cite{Seiferle2019}\\
$\Gamma_\gamma$ & Radiative decay rate & $10^{-4}$ Hz & Estimated from theory \cite{Minkov2019} \\
$\alpha_\text{ic}$ & IC coefficient & $10^9$ & Based on $\Gamma_\gamma$ and the IC lifetime \cite{Seiferle2} \\ 
$N_0$ & Irradiated atoms & $1.6\cdot10^{13}$ & For $10$ nm thickness and 0.3 mm diameter ThO$_2$\\
\noalign{\smallskip}\hline
\end{tabular}
\label{table_surface}
\end{center}
\end{table*}

\subsection{The detection system}
\begin{figure}[t]
\begin{center}
\includegraphics[totalheight=4cm]{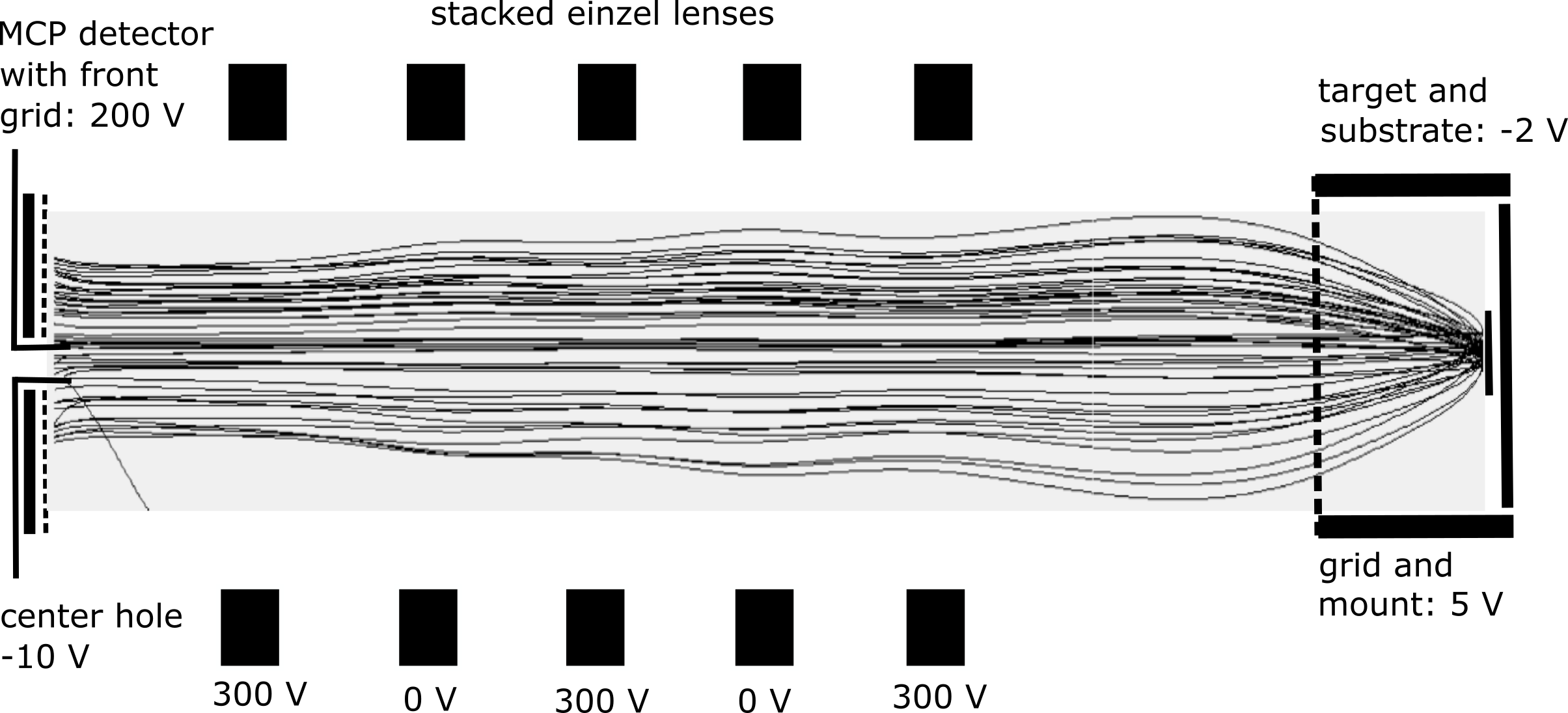}
\caption{Schematic drawing of the detection system proposed for low-energy IC-electron detection. Electrons emitted at the $^{229}$Th target are guided by electrostatic lenses to an MCP detector with a center hole. SIMION \cite{Dahl} simulations were performed showing that guiding efficiencies of more than 80\% for electrons between 0.1 and 2.0 eV can easily be achieved. Potentials used for the simulations are shown in the figure together with 50 electron trajectories obtained for electrons with 0.5~eV kinetic energy. The total distance between target and detector is about 10~cm.}
\label{detection}
\end{center}
\end{figure}

It is proposed to use an MCP detector \cite{Wiza} (Hamamatsu, type F2223-21SH, 27 mm diameter active area) for the detection of the low-energy IC electrons emitted in the isomeric decay. A center hole is foreseen in order to allow for $^{229}$Th target irradiation in a simple geometry. High detection efficiencies of about 50\% for electrons are achieved when post-accelerated to about 300~eV kinetic energy \cite{Goruganthu}. The detector should be placed in some distance to the target in order to reduce the radioactive background in the form of high-energy particles. For this reason a stack of electrostatic einzel-lenses could be used to guide only electrons of low kinetic energy to the MCP detector. High energy $\alpha$ or $\beta$ particles do not follow the electric fields and are implanted into the chamber walls. SIMION \cite{Dahl} simulations were performed in order to optimize the electric potentials of the guiding fields and estimate the guiding efficiency. A sketch of the electrode configuration together with the voltages and electron trajectories obtained from the SIMION simulations are shown in Fig.~\ref{detection}. The distance between the target and the detector is about 10~cm. At the given potentials, high guiding efficiencies of more than 80\% were achieved for electrons of kinetic energies between 0.1 and 2.0~eV. The fraction of the unit sphere that is covered by the MCP detector at 10 cm distance amounts to about 0.45\%. Assuming a total target activity of 2~kBq, about 9 high-energy $\alpha$ and $\beta$ particles are expected to hit the detector per second. This is negligible compared to the expected background induced by low-energy electrons emitted as a byproduct in $\alpha$ decay (see next sub-section).\\[0.2cm] 
A grid is positioned in front of the target, which allows the application of attractive electric fields in the time window meant for electron detection, but repelling electric fields during the time of laser irradiation. This will prevent photoelectrons emitted at the target from reaching the detector, that would otherwise lead to detector saturation \cite{Wense4}. The total efficiency of the detection system is estimated to be 40\%.\\[0.2cm]
While the here proposed MCP detection scheme appears to be the most straight-forward approach, one should also consider different ways for IC electron detection in the future, that do not require the electrons to leave the target material. One could imagine, for example, detection schemes used in charge-coupled devices (CCDs) \cite{Boyle} or in transition-edge detectors like superconducting nanowire single-photon detectors (SNSPDs) \cite{Natarajan}. In this case not the amount of electrons that leaves the target surface, but instead the inverse amount would contribute the signal, thereby drastically increasing the expected count rate.

\subsection{Quantitative evaluation}
\begin{figure}[t]
\begin{center}
\includegraphics[totalheight=4.5cm]{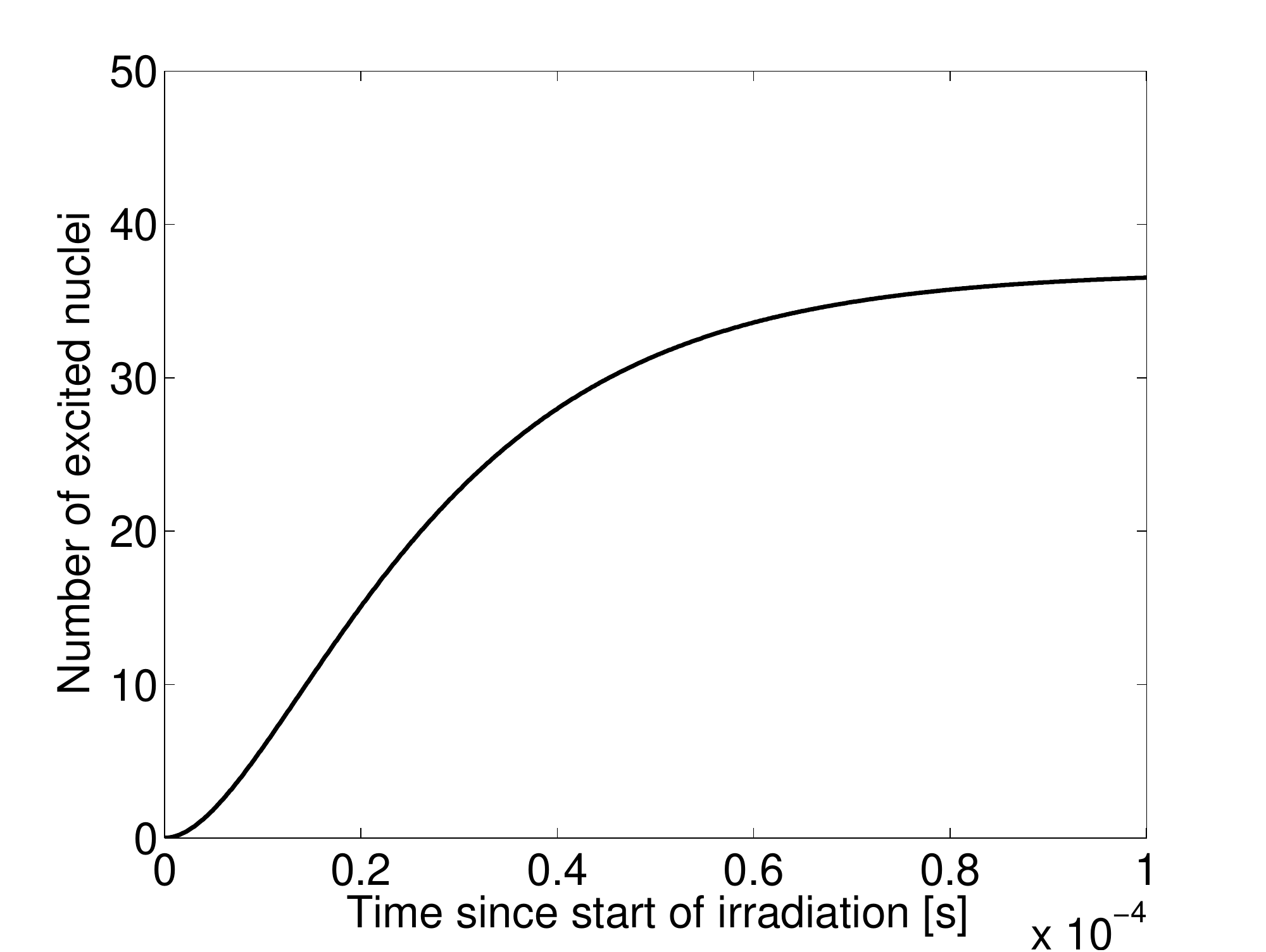}
\caption{\small Expected number of excited nuclei as a function of time if $1.6\cdot10^{13}$ $^{229}$Th atoms in a ThO$_2$ surface are irradiated in parallel. The calculation was performed based on Eq.~(\ref{Torrey}) using $C_{ge}=1$ and values for variables listed in Tab.~\ref{table_surface}. A loss factor of $\eta_\text{abs}=0.63$ due to VUV absorption within the target was already taken into account.}
\label{excnucl_concept1}
\end{center}
\end{figure}

The number of nuclear excitations as a function of time is shown in Fig.~\ref{excnucl_concept1}. It was calculated based on Eq.~(\ref{Torrey}) using the $^{229}$Th target- and laser parameters listed in Tab.~\ref{table_surface} and multiplied with the loss factor due to VUV surface absorption of $\eta_\text{abs}=0.63$. About $36.5$ nuclei can be expected to be excited into the isomeric state during each laser gate pulse of 100~$\mu$s duration. This value is close to the maximum as obtained in equilibrium of $37$ excited nuclei. These nuclei will decay to the nuclear ground state with about 10~$\mu$s lifetime via the emission of low-energy internal conversion electrons \cite{Seiferle2}. Given the gate frequency of 5~kHz, up to $1.8\cdot10^5$ nuclear excitations are expected to occur per second in case of resonance. It was shown that time gating is sufficient for the exclusion of photoelectrons as a background, if a time window from 5 to 10 $\mu$s after the end of each laser pulse is chosen for IC electron detection \cite{Wense4}. However, low-energy electrons being continuously emitted from the radioactive $^{229}$ThO$_2$ target in the form of, e.g., shake-off electrons from shell-reorganization after $\alpha$ decay \cite{Wandkowsky}, Auger electrons, conversion cascades, etc., have to be considered as the main background. Due to the complexity of these processes, a satisfactory quantitative estimate of this background will require a detailed experimental investigation. We here roughly estimate that, on average, 2 low-energy electrons per $\alpha$ decay are emitted from the target \cite{Wandkowsky}. First preliminary experimental investigations indicate that this estimate is indeed valid. Of the assumed target activity of 1.7~kBq, 1.3~kBq originates from $\alpha$ decay, resulting in an expected emission of about $2.6\cdot10^3$ low-energy electrons per second. On average, this corresponds to 0.013 emitted electrons in the 5~$\mu$s detection window. The number of expected isomeric decays in the same time window amounts to $0.24\cdot N_\text{exc}\approx8.9$. Even when assuming a factor of 10 of IC electron losses in the target, the estimated signal-to-background ratio is about $68:1$ in case of resonance. Here the guiding and detection efficiency for IC and background electrons was assumed to be identical. Note that the signal-to-background ratio improves when the source ages, as the $^{228}$Th intrinsic activity will fade away with 1.9~years of half-life. The absolute number of detectable IC electrons amounts to $\sim0.53$ in the 5~$\mu$s detection window per laser gate pulse. This includes a 15\% probability for the electrons to leave the target and a 40\% total efficiency of the detection system. Considering $5000$ gate pulses per second, about $2.6\cdot10^3$ isomeric electrons could be detected per second in resonance in the $5$~$\mu$s detection window.\\[0.2cm] 
During the search for the isomeric excitation, the frequency comb has to be tuned. As the scan is performed with all $1.2\cdot10^5$ comb modes in parallel, it is sufficient to bridge the mode spacing of $77$~MHz of two consecutive comb modes in order to probe the total bandwidth of the frequency comb of 0.038~eV (9.2 THz). Considering the IC-broadened nuclear transition linewidth of 15.9~kHz, about $5000$ scan steps will therefore be required. Assuming that 100 gate pulses per scan step are used, the total time required for scanning of the 0.038~eV energy range amounts to 100 seconds. In order to cover the 0.34~eV energy range corresponding to the 1~$\sigma$ uncertainty interval of the currently best energy constraint of $8.28\pm0.17$~eV \cite{Seiferle2019}, 9 of such scans would have to be performed leading to a minimum total scanning time of 900~s. The main parameters of the proposed method are shown in Tab.~\ref{concept1_main}.\\[0.2cm]

\begin{table}[t]
\begin{center}
\caption{Main parameters for ThO$_2$ target irradiation.}
\begin{tabular}{lc}
\hline\noalign{\smallskip}
Description & Value \\
\noalign{\smallskip}\hline\noalign{\smallskip}
Number of excited nuclei per pulse & 37  \\
Number of gate pulses per second & 5000 \\
Time per scan step & 0.02 s \\
Number of excited nuclei per scan step & $3.7\cdot10^3$ \\
Number of scan steps for 0.34 eV interval & $4.5\cdot10^4$ \\
Time required to scan 0.34 eV & 15 min  \\
Signal-to-background ratio & $\sim$68:1  \\
\noalign{\smallskip}\hline
\end{tabular}
\label{concept1_main}
\end{center}
\end{table}

\section{Line splitting and broadening effects in a solid-state environment}
\label{Splitting}
As is known from Mössbauer spectroscopy, several line splitting and broadening effects have to be considered in a solid-state environment. A detailed discussion of these effects for the $^{229}$Th nuclear transition can be found in Refs.~\cite{Rellergert,Kazakov}. Here the discussion will be adapted to a pure $^{229}$ThO$_2$ environment. ThO$_2$ is chosen as a target material as it is chemically stable and provides a cubic lattice structure resulting in no measurable quadrupole splitting.

\subsection{Nuclear hyperfine splitting}
\begin{figure}[t]
\begin{center}
\includegraphics[width=8cm]{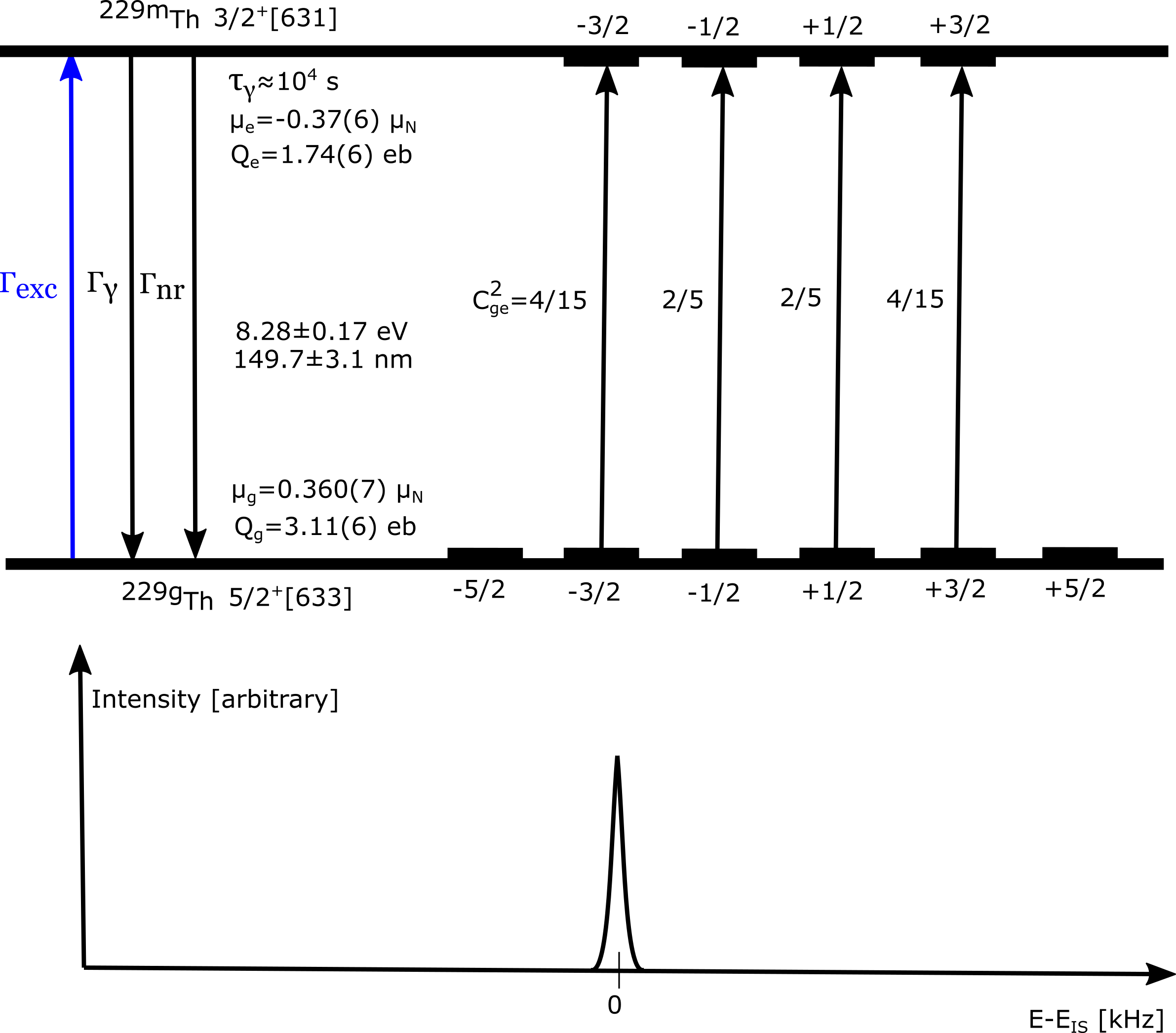}
\caption{Nuclear magnetic sub-states of $^{229}$Th embedded in a ThO$_2$ environment. ThO$_2$ exhibits a cubic lattice structure, therefore no electric quadrupole shift will occur. In addition, no unpaired electron spins are present, therefore no intrinsic Zeeman splitting caused by the electronic environment will arise. Accordingly, the nuclear spectrum consists of a single line as long as no other chemical bondings are present and no external magnetic field is applied.}
\label{singleline}
\end{center}
\end{figure}

A hyperfine splitting of the nuclear transition occurs due to the coupling of the spins of the nuclear ground- and excited state to the electronic environment. This splitting consists of two parts: (1) a quadrupole splitting that arises from the nuclear quadrupole moment in an electric field gradient that is generated at the point of the nucleus by the surrounding electrons and (2) a Zeemann splitting, which originates from a magnetic field that can either be intrinsic to the sample or externally applied. Quantitatively, the line splitting is described by the equation \cite{Greenwood}
\begin{equation}
\label{hyperfine}
\Delta E_\text{HFS}=g_I\mu_Nm_I B+\frac{Q^{(s)} V_{zz}}{4} \frac{3m_I^2-I(I+1)}{I(2I-1)}.
\end{equation}
Here $B$ denotes the magnetic field and $V_{zz}$ the second derivative of the electric potential at the site of the nucleus. $g_I$ is the nuclear Landé g-factor, which is related to the nuclear magnetic moment via $\mu_I=g_I\mu_NI$, $\mu_N$ is the nuclear magneton and $Q^{(s)}$ denotes the spectroscopic quadrupole moment of the nucleus. Further, $I$ and $m_I$ are the spin and magnetic quantum number, respectively, corresponding to the nuclear sub-state under consideration, where $m_I$ takes the $2I+1$ values from $-I$ to $+I$.\\[0.2cm]
Thorium-dioxide ($^{229}$ThO$_2$) provides a cubic crystal structure. For this reason $V_{zz}\approx 0$ holds and no significant quadrupole splitting is expected to be observable at the point of the $^{229}$Th nucleus \cite{Knop}. Experimentally, Mössbauer spectroscopy that was performed with the 49.4~keV $\gamma$-ray of $^{232}$Th in a $^{232}$ThO$_2$ crystal structure supports this expectation \cite{Hershkowitz}. However, the half-life of the 49.4~keV nuclear excited state of $^{232}$Th is only 345~ps, resulting in a natural linewidth of about 320~MHz, which could easily cover a small quadrupole splitting. For this reason, a comparison to the long-lived Mössbauer state at 93.3~keV in $^{67}$Zn may provide more useful information. This Mössbauer state has a half-life of 9~$\mu$s, similar to the 7~$\mu$s observed half-life of $^{229\text{m}}$Th under IC decay \cite{Seiferle2}. The natural linewidth of the $^{67}$Zn Mössbauer transition is therefore with 12.4~kHz comparable to the one of $^{229\text{m}}$Th. Also for $^{67}$Zn no quadrupole splitting could be observed when a cubic crystal structure ($^{67}$ZnS) was used for Mössbauer spectroscopy \cite{Forster}. This gives a clear indication, that, despite the narrow expected linewidth of the $^{229\text{m}}$Th nuclear transition of 15.9~kHz, no quadrupole splitting of the nuclear hyperfine levels will be observable.\\[0.2cm]
In addition, ThO$_2$ is paramagnetic and does not exhibit unpaired electron spins, therefore we have $B\approx0$ and no magnetic hyperfine splitting will arise due to target intrinsic magnetic fields potentially generated by the electronic environment. It was discussed in Refs.~\cite{Rellergert,Kazakov}, that a small magnetic field will be generated due to the surrounding nuclear spins. As the nuclear spins are randomly oriented, the corresponding Zeeman splitting will lead to a homogeneous line broadening of the transition, which is assumed to be the dominant line broadening effect in a solid-state environment. The corresponding broadening was estimated to be between $2\pi\cdot1$ and $2\pi\cdot10$~kHz in Ref.~\cite{Rellergert} and to about $2\pi\cdot150$~Hz in Ref.~\cite{Kazakov}. In any case the broadening effect is smaller than the IC-broadened natural $^{229\text{m}}$Th transition linewidth of $2\pi\cdot15.9$~kHz and will therefore not significantly affect the considered concept. This is in agreement with the fact that it was possible to experimentally resolve the natural transition linewidth ($2\pi\cdot 12.4$~kHz) of the $^{67}$Zn Mössbauer line \cite{Potzel}.\\[0.2cm]
In conclusion, the Mössbauer spectrum of $^{229}$ThO$_2$ will consist of a single line as shown in Fig.~\ref{singleline}, if no other crystal structures are present and no external magnetic field is applied. Importantly, a chemical (or isomer) shift, as often discussed in the context of Mössbauer spectroscopy, corresponds to an absolute shift of the excitation energy due to the interaction of the nucleus with its electron cloud \cite{Shenoy}. It is not related to a line splitting. Although potentially being large (on the order of 100~MHz) the shift is constant for a constant chemical environment and does therefore not affect the proposed experimental scheme.

\subsection{Line broadening effects}
It is a central feature of Mössbauer spectroscopy, that no first order Doppler broadening of the transition line as well as energy shifts due to the $\gamma$-ray recoil have to be considered, as the recoil momentum is transferred to the entire crystal if the atoms are confined in the Lamb-Dicke regime \cite{Moessbauer,Gibb,Frauenfelder}. The physical reason is, that in a crystal lattice structure the atoms are strongly bound and are vibrating fast with a small amplitude around their center position. If this amplitude is smaller than the wavelength of the transition the Doppler effect averages out, resulting in a Doppler-free linewidth (the effect is known as the Dicke effect) \cite{Frauenfelder,Dicke,Lamb}. The fraction of atoms that exhibit a Doppler-free linewidth is given by the Debye-Waller factor $f=\exp(-E_\gamma^2\langle x^2\rangle/(\hbar c)^2)$, where $\langle x^2\rangle$ denotes the mean-square vibrational amplitude of the nucleus. Usually, $\langle x^2\rangle$ is calculated in the frame of the Debye model, leading to \cite{Gibb,Frauenfelder}
\begin{equation}
f=\exp\left(-\frac{6E_R}{k\Theta_D}\left(\frac{1}{4}+\left(\frac{T}{\Theta_D}\right)^2\int_0^{\Theta_D/T}\frac{xdx}{e^x-1}\right)\right).
\end{equation}
Here $E_R=E_\gamma^2/(2Mc^2)$ denotes the nuclear recoil energy, $k$ is the Boltzmann constant, $\Theta_D$ the Debye temperature and $T$ the temperature of the crystal. For high-energy $\gamma$ rays the Debye-Waller factor is only large, if the crystal temperature is significantly below $\Theta_D$, which may require cooling to cryogenic temperatures. In the special case of $^{229\text{m}}$Th, however, $E_\gamma$ is sufficiently small so that large values for $f$ are obtained over a wide temperature range. Reported literature values for the Debye temperature of ThO$_2$ vary significantly (see, e.g., Refs.~\cite{Willis,Ali,Serizawa,Lu,Kelly}), however, the Debye-Waller factor for $^{229\text{m}}$Th remains largely unaffected. For an assumed Debye temperature of $\Theta_D=400$~K, a Debye Waller factor close to one is obtained at room temperature.\\[0.2cm]
A second order Doppler broadening will be present due to vibrations of the $^{229}$Th nuclei at finite temperature. This effect was estimated in Ref.~\cite{Kazakov} to about $2\pi\cdot T$~Hz, where $T$ denotes the temperature in Kelvin. At room temperature the broadening will amount to less than 2~kHz, which is significantly below the IC-broadened nuclear transition linewidth. For this reason no temperature related second order Doppler broadening is expected to be detectable even at room temperature and the experiment does not require a cryogenic environment.\\[0.2cm]
However, two other line-broadening effects require a discussion: homogeneous broadening from rapid fluctuations of the electronic environment and inhomogeneous broadening from different chemical environments in the target. An inhomogeneous broadening may arise, if different chemical bondings in the $^{229}$ThO$_2$ matrix are present, as this would potentially lead to different hyperfine splittings \cite{Rancourt}. This can be avoided even in thin surfaces by using a pure target material. A conversion-electron Mössbauer spectroscopy (CEMS) experiment using a 25~nm thin $^{57}$Fe absorber was performed in Ref.~\cite{Shigematsu}. The observed Mössbauer spectrum showed the expected unbroadened hyperfine splitting. A small absorber thickness was required, as the inelastic mean-free path of the observed electrons is only about 10~nm, comparable to the one expected for $^{229\text{m}}$Th.\\[0.2cm]
 A further homogeneous line broadening will be present, as the electronic environment can usually not be treated as static and the hyperfine structure will in general be subject to time-dependent fluctuations \cite{Rancourt}. Fluctuations may arise in the temperature and as the electron density at the site of the nucleus is temperature dependent, this will lead to fluctuations in the chemical shift. The observed energy change caused by temperature is typically on the order of 10 kHz/K (see Ref.~\cite{Rellergert} and references therein). It was shown in Sec.~\ref{sec2} that, despite the laser irradiation of the target, temperature fluctuations can be kept significantly below 1~K and the resulting homogeneous broadening is therefore expected to be negligible compared to the IC-broadened natural linewidth of the transition.\\[0.2cm]
According to the first term of Eq.~(\ref{hyperfine}), fluctuations of the magnetic field at the site of the nucleus will lead to a homogeneous line-broadening. As the intrinsic magnetic fields of the target are only of minor concern, such broadening effects can be kept at a minimum by stabilizing the external magnetic field of the environment. A stabilization to better than 1~mT is required in order to keep any potential line broadening to below 10~kHz. Experimentally this can easily be achieved. Due to the cubic lattice structure, no temperature-dependent line-broadening effects that could potentially be caused by the coupling of a fluctuating electric field gradient to the nuclear quadrupole moment have to be considered.\\[0.2cm]
The above quantitative discussion shows, that during laser-based CEMS of $^{229\text{m}}$Th in a thin $^{229}$ThO$_2$ layer, only a single transition line with the natural linewidth of 15.9~kHz has to be expected. This situation changes if (1) a magnetic field is externally applied or (2) a different (non-cubic) crystal structure is used. Both situations will be quantitatively discussed in the following.

\subsection{External magnetic fields}
\begin{figure}[t]
\begin{center}
\includegraphics[width=8cm]{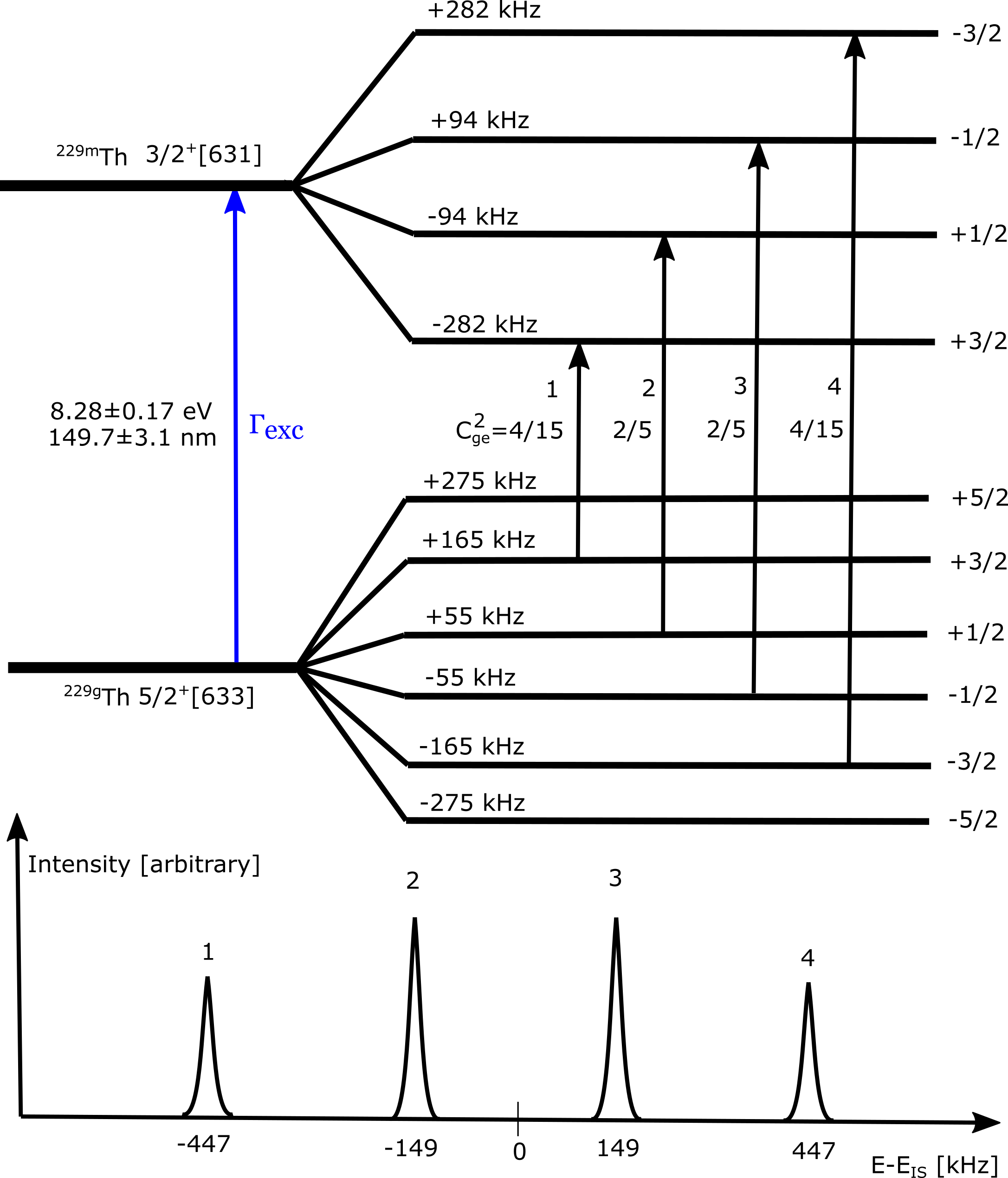}
\caption{Expected Zeeman splitting of the $^{229\text{m}}$Th nuclear transition in a CEMS experiment in case of a pure $^{229}$ThO$_2$ absorber and an externally applied 100~mT constant magnetic field. It is assumed that linearly polarized laser light is used for excitation. The excitation probability scales proportional to the Clebsch-Gordan coefficients squared ($C_{ge}^2$), therefore the central lines are by a factor of 1.5 more intense than the outer lines.}
\label{magnsplitting}
\end{center}
\end{figure}

According to the Zeeman effect described by the first term of Eq.~(\ref{hyperfine}), a line splitting will arise as soon as an external magnetic field is applied. For the nuclear ground-state ($I_g=5/2$) a magnetic moment of $\mu_g=0.360(7)\mu_N$ was experimentally determined \cite{Safronova2013}. For the nuclear excited-state ($I_e=3/2$) the corresponding value is $\mu_e=-0.37(6)\mu_N$ \cite{Thielking}. This allows one to calculate the expected line splitting. The splitting of the magnetic sub-states as well as the transitions that can be driven with linearly polarized laser light are shown in Fig.~\ref{magnsplitting} together with the expected Mössbauer spectrum under the assumption that the nuclei are exposed to a constant magnetic field of 100~mT.
In this case the splitting between the two outermost lines is determined to be about 900~kHz. As the splitting scales proportional to $B$, no line splitting will be observed as long as the magnetic field is kept below 1~mT. This value also corresponds to the stability of the magnetic field required to suppress any line broadening due to the Zeeman effect. As soon as the isomeric state has been successfully excited, magnetic fields can be applied in order to verify the signal origin. Importantly, if no magnetic field is applied, all four lines will be degenerate and their intensities will add up. Therefore the actual number of nuclear excitations in equilibrium should be by a factor of 1.3 larger than the one stated in Tab.~\ref{concept1_main}, where only one pair of magnetic sub-states was considered, however with $C_{ge}=1$.   

\section{\label{clock} A new solid-state nuclear clock concept}
\noindent The development of a nuclear optical clock based on individual $^{229}$Th$^{3+}$ ions in a Paul trap has been extensively discussed in literature \cite{Peik,Campbell,Peik2}. Such a clock is expected to achieve an extraordinary high accuracy approaching $10^{-19}$ \cite{Campbell}. Although expected to be less accurate \cite{Rellergert}, also a nuclear clock that makes use of a large number of $^{229}$Th ions embedded in a crystal-lattice environment has attracted significant attention, as it may pose a practical tool for time measurement \cite{Peik,Rellergert,Kazakov}. Here, a new solid-state nuclear clock concept is proposed, which is based on conversion-electron Mössbauer spectroscopy (CEMS) \cite{Shigematsu}. For this purpose a narrow-band laser is stabilized to the nuclear transition of $^{229}$Th deposited as a thin layer of $^{229}$ThO$_2$ onto a substrate surface. Successful nuclear excitation is probed via detection of the IC electrons emitted shortly after the laser pulse, just as proposed in Sec.~\ref{sec2}.
It will be shown in the following, that this concept might be advantageous compared to the crystal-lattice clock approach \cite{Rellergert,Kazakov}.\\[0.2cm]
Clock performance is generally expressed via two parameters: accuracy and stability. The stability of a clock corresponds to the statistical uncertainty of a frequency measurement and depends on the measurement time. Opposed to that, the accuracy (or more precisely: the systematic frequency uncertainty) of a clock also takes all systematic uncertainties of a frequency comparison between different clocks into consideration.
In order to achieve a high clock performance, it is important that the statistical uncertainty approaches the systematic uncertainty within short averaging times. On the other hand, any improvement of the statistical uncertainty of the frequency measurement significantly below the value of the systematic frequency uncertainty will not improve the clock performance, as this will be limited by other systematic effects.\\[0.2cm]
It was shown in Ref.~\cite{Kazakov} that the stability of a $^{229}$Th-based solid-state nuclear clock can be estimated for the realistic case of low laser intensities and short interrogation times ($T<1/\Delta\omega_0$) via a shot-noise limited Allan deviation of\footnote{The equation is obtained by inserting Eq.~(37) of Ref.~\cite{Kazakov} into Eq.~(48) of the same paper and using that by definition $t=4T$, where $t$ denotes the time for a complete interrogation cycle used in Ref.~\cite{Kazakov}.}
\begin{equation}
\label{sigma}
\sigma_y(\tau)\approx\frac{1}{\omega_0 T}\frac{\tilde{\Gamma}}{\Gamma}\frac{1}{\sqrt{\Gamma_\text{exc}N_\text{eff}\tau}}.
\end{equation}
Here $T$ denotes the interrogation time, which should not exceed the isomeric lifetime, $\tilde{\Gamma}$ is the decay rate of the coherences given in Eq.~(\ref{Gammatilde}) and $\Gamma=\Gamma_\gamma+\Gamma_\text{nr}$ the total decay rate of the transition, including non-radiative decay channels. $N_\text{eff}$ denotes the effective number of irradiated nuclei defined as $N_\text{eff}\approx\Gamma_\gamma/\Gamma \cdot k S/(4\pi)\cdot N_0$, with $k$ being the quantum efficiency, $S$ the effective solid angle covered by the detector and $N_0$ the actual number of irradiated nuclei. $\tau$ is the averaging time, which has to be equal or larger than the time $T$ chosen for interrogation and $\Gamma_\text{exc}$ is the nuclear excitation rate, which is estimated for the considered case of low laser intensities based on Eq.~(\ref{rate}).\\[0.2cm]
In the following, the crystal-lattice approach as proposed in Refs.~\cite{Peik,Rellergert} is discussed. Assuming purely radiative decay ($\Gamma=\Gamma_\gamma$) and an interrogation time of $T=10^4$~s, corresponding to the expected radiative lifetime, as well as a decay rate of the coherences of $\tilde{\Gamma}\approx 1$~kHz due to coupling to the crystal-lattice environment (see Ref.~\cite{Kazakov} for details), a straight forward calculation reveals a high stability of $\sigma_y(\tau)\approx 6.0\cdot10^{-17}/\sqrt{\tau}$ for a solid state nuclear optical clock \cite{Kazakov}. Here, values of $N_0=10^{13}$, $k=0.1$, $S=4\pi/10$ and $I=1.4\cdot10^{-5}$~W/cm$^2$ (corresponding to 10~nW of laser power irradiating a crystal of 0.3~mm diameter) were used. Thus, after the shortest useful averaging time of $\tau=10^4$~s, a stability in the $10^{-19}$ range appears to be achievable. However, the systematic frequency uncertainty of the clock is expected to be limited by other effects, most importantly temperature shifts of the crystal, to a value of $2\cdot10^{-16}$ \cite{Rellergert}. For this reason only limited advantage will be gained from the high stability as long as other systematic shifts will not be sufficiently well controlled.\\[0.2cm]
In the following, the same stability estimate is applied to a new nuclear clock concept, which is based on the observation of internal conversion electrons following the laser irradiation of $^{229}$ThO$_2$ deposited on a surface. In this case, the nuclear decay rate will be dominated by the short IC lifetime of about $10$~$\mu$s \cite{Seiferle2}: $\Gamma\approx\Gamma_\text{ic}\approx10^{5}$~Hz. Further, as the laser used for irradiation is narrowband, the decay rate of the coherences will be dominated by the nuclear decay rate: $\tilde{\Gamma}\approx\Gamma/2\approx\Gamma_\text{ic}/2$. Inserting these values, together with a short interrogation time of $T=10$~$\mu$s, into Eq.~(\ref{sigma}), a stability of $\sigma_y(\tau)\approx9.5\cdot10^{-15}/\sqrt{\tau}$ is achieved. Here it was assumed that $N_0=10^{13}$ atoms are irradiated with a laser of intensity $I=1.4\cdot10^{-5}$~W/cm$^2$. Further, $N_\text{eff}=k S/(4\pi)$ was used, taking into account that no competing decay channels exist, with $k=0.1$ as before, but $S=4\pi/2$, as the emitted electrons can be attracted towards the detector. Obviously, the calculated stability performance is lower than the best that could be achieved with a crystal-lattice nuclear clock. Importantly, however, after about $2.5\cdot10^3$~s of averaging the stability has surpassed the expected systematic frequency uncertainty, for which a value of $2\cdot10^{-16}$, like for the crystal-lattice approach, is assumed. Of course there is the option to choose shorter interrogation times also in the crystal-lattice clock approach. This will, however, significantly affect the clock's stability performance, as it scales proportional to $1/T$ (e.g., using $T=1000$~s interrogation time, a stability of $6.0\cdot10^{-15}/\sqrt{\tau}$ is obtained, comparable to the expected stability of an IC-based nuclear clock).\\[0.2cm]
From the above considerations it is evident that the stabilities of both, a crystal-lattice nuclear clock as well as an IC-based nuclear clock, should be sufficient to approach a statistical uncertainty of the frequency measurement of about $10^{-16}$ after a few thousand seconds of averaging. Longer measurement times are not expected to improve clock performance, which will most likely be limited to a level of $\sim10^{-16}$ by systematic frequency uncertainties \cite{Rellergert}. However, it may be seen as a big advantage of the IC-based nuclear clock approach that no competing decay channels exist next to internal conversion, which has already been experimentally observed and is known to be the dominant decay channel for the considered experimental conditions. Opposed to that, the crystal-lattice nuclear clock makes important use of the $^{229\text{m}}$Th $\gamma$ decay, which requires suppression of the IC decay channel by nine orders of magnitude in a solid-state environment. Even when achieved, non-radiative decay via bound internal conversion (BIC) and electronic bridge (EB) mechanisms might be present, reducing the isomeric lifetime and the fraction of observable $\gamma$ decays. 

\section{Laser spectroscopy of $^{229}$Th$^{3+}$ ions in a Paul trap}
\label{concept2sec}
\noindent Most of the $^{229}$Th-related research in the past decade has focused on the problem of constraining the isomeric energy value by about a factor of 10 to 50~meV uncertainty, corresponding to the total bandwidth of a frequency comb in the VUV region. However, the problem of improving the energy uncertainty down to a few 100~Hz, as required for the development of single-ion nuclear clock, has not been addressed. Importantly, laser excitation of $^{229}$Th ions or atoms in a solid-state environment does not solve this problem, as the nuclear hyperfine structure splitting is expected to deviate by several 100~MHz from the single-ion case. In the following this problem will be addressed. As soon as the isomeric energy has been constrained to sufficient precision, a single-ion nuclear clock can be build, which requires to drive nuclear Rabi oscillations. The potential for driving nuclear Rabi oscillations will be discussed in Sec.~\ref{concept3sec}.\\[0.2cm]

\subsection{Excitation of multiple trapped $^{229}$Th$^{3+}$ ions}
\label{multipleions}
It is assumed that the laser system already presented in Sec.~\ref{sec2} is used to irradiate 10 laser-cooled $^{229}$Th$^{3+}$ ions in a Paul trap. Laser cooling could be achieved either directly \cite{Campbell2011} or via sympathetic cooling \cite{GrootBerning2019}. Further, it is assumed that the laser light is focused to a spot of 3 $\mu$m diameter in order to irradiate the chain of ions and that the energy has been constrained by a different method, e.g., the one described in Sec.~\ref{sec2}, to a value that corresponds to the total bandwidth of the frequency comb used for excitation (0.038~eV in the considered case)\footnote{Importantly, any improvement of the energy uncertainty to a value below the total bandwidth of the frequency comb does not lead to a shortening of the scanning time as the frequency spacing between two consecutive comb modes will always have to be scanned.}. In this concept, the successful laser excitation is probed via exploiting the change of the hyperfine structure of the electronic shell, induced by the different spins of nuclear ground- and excited state, which is known as the double-resonance method \cite{Peik}. In the following, the expected number of excited nuclei as a function of time will be calculated based on Eq.~(\ref{Torrey}).\\[0.2cm]
In $^{229}$Th$^{3+}$ ions the internal conversion decay channel is energetically suppressed and $\alpha_\text{ic}=0$ holds. However, a potential electronic bridge (EB) decay channel might exist and lead to an enhancement of the total decay rate according to $\Gamma=\Gamma_\gamma+\Gamma_\text{eb}$, where $\Gamma_\text{eb}$ denotes the decay rate related to the EB process and replaces $\Gamma_\text{ic}$ of the previous considerations. In the EB decay the nucleus couples to the electronic shell, exciting an electron into a virtual shell state, which immediately decays under photon emission \cite{Strizhov}. For $^{229}$Th$^{3+}$, the EB coefficient $\alpha_\text{eb}=\Gamma_\text{eb}/\Gamma_\gamma$ was estimated to be between 0.01 and 0.1 \cite{Porsev3} for the electronic ground state and between $20$ \cite{Porsev3} and about $50$ \cite{Mueller2017} for the metastable 7s$^2$S$_{1/2}$ electronic excited state. For the following calculations $\alpha_\text{eb}=50$ is assumed, resulting in an isomeric lifetime which is reduced by a factor of $50$ to $\tau_\text{eb}\approx 200$~s compared to the radiative lifetime $\tau_\gamma=1/\Gamma_\gamma\approx10^4$~s. Experimentally, a lower limit of 60~s for the isomeric lifetime in $^{229}$Th$^{2+}$ ions was observed \cite{Wense1}. The parameters used to calculate the nuclear excitation probability via Eq.~(\ref{Torrey}) are listed in Tab.~\ref{input_concept2}. As the laser intensity is sufficiently strong to enter the power-broadened regime, the result is not significantly affected by the EB decay channel. Following Eq.~(\ref{Torrey}) this is correct as long as $\tilde{\Gamma}\Gamma\ll\Omega_{eg}^2$ holds. In the considered case one has $\Gamma\ll\Gamma_L$, which results in combination with Eq.~(\ref{Gammatilde}) in the condition $\Gamma\ll 2\Omega_{eg}^2/\Gamma_L\approx 0.22$~Hz. For this reason, the concept is not affected as long as $\alpha_\text{eb}\ll 2200$ holds. However, EB coefficients significantly larger than $100$ are experimentally excluded \cite{Wense1}.
\begin{table}[t]
\begin{center}
\caption{List of input values used in Eq.~(\ref{Torrey}) for calculation of the number of nuclear excitations in a chain of $10$ trapped $^{229}$Th$^{3+}$ ions.}
\begin{tabular}{ccl}
\hline\noalign{\smallskip}
Variable  & Value & Comment \\
\noalign{\smallskip}\hline\noalign{\smallskip}
$I$  & $0.14$ W/cm$^{2}$ & Single-mode focused to $\varnothing$ $3$~$\mu$m\\
$\Gamma_L$  & $2\pi\cdot490$ Hz & Larger than nuclear linewidth\\
$\omega_0$  & $2\pi\cdot2.0$ PHz & Corresponding to 8.3 eV energy\\
$\Gamma_\gamma$  & $10^{-4}$ Hz & Estimated from theory \\
$\alpha_\text{eb}$  & $50$ & EB decay for Th$^{3+}$ ions \cite{Mueller2017} \\ 
$N_0$  & $10$ & Multiple laser-cooled ions\\
\noalign{\smallskip}\hline
\end{tabular}
\label{input_concept2}
\end{center}
\end{table}

\begin{figure}[t]
\begin{center}
\includegraphics[totalheight=4.5cm]{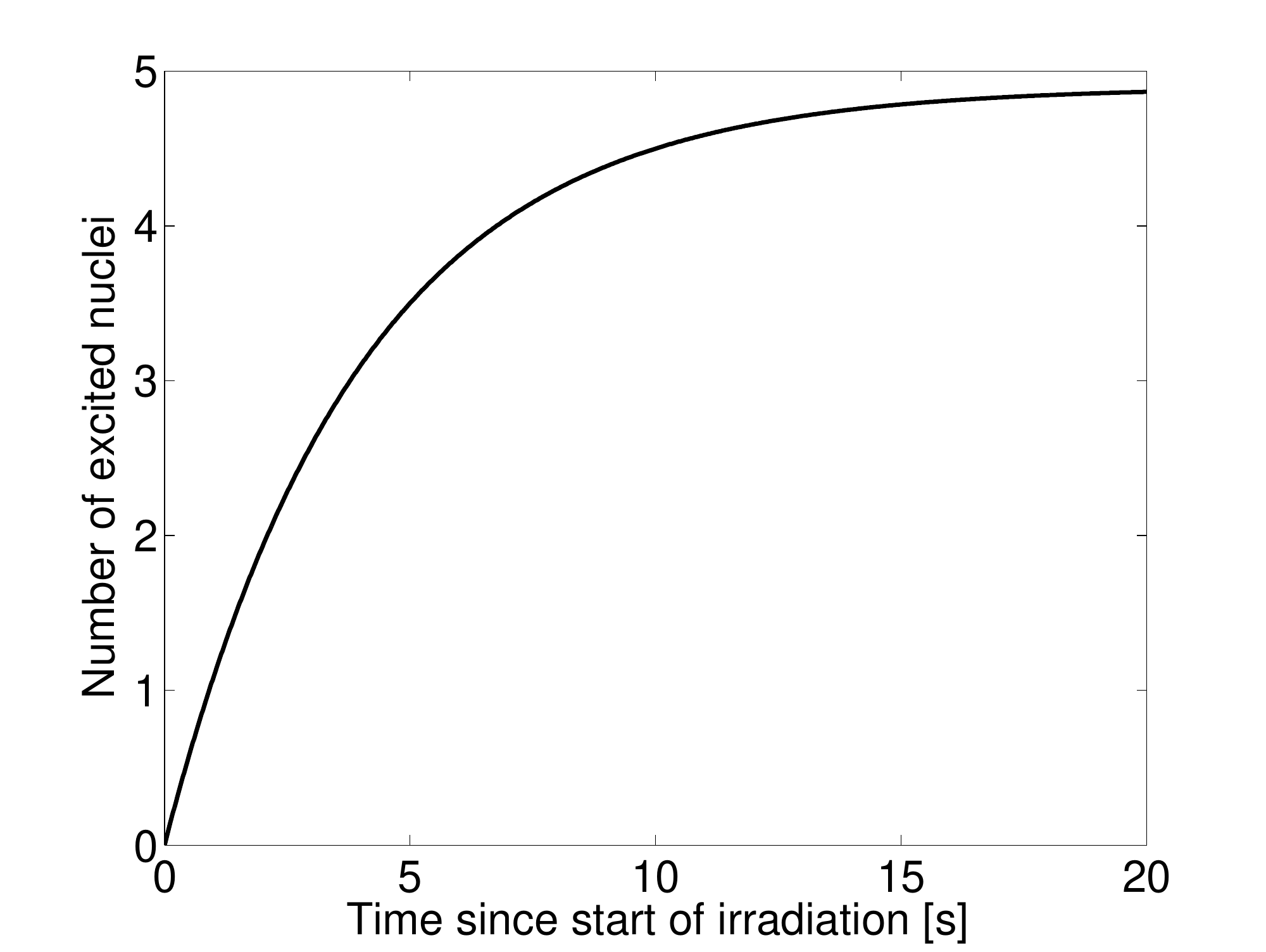}
\caption{\small Expected number of excited nuclei as a function of time if 10 $^{229}$Th$^{3+}$ ions in a Paul trap are irradiated in parallel. The curve is generated based on Eq.~(\ref{Torrey}) with $C_{ge}=1$, making use of the input values listed in Tab.~\ref{input_concept2}.}
\label{excnucl_concept2}
\end{center}
\end{figure}

\noindent The result of the calculation is shown in Fig.~\ref{excnucl_concept2}. Here a factor $C_{ge}=1$ was used, making no assumption on the nuclear magnetic sub-system. In a real experiment, a particular sub-system has to be chosen, where each possible transition is defined by the quantum numbers of the ground and excited nuclear state corresponding to the electronic angular momentum $J$, the nuclear spin $I$, the total angular momentum $F=I+J$, as well as the magnetic quantum number $m$. These define the multipolarity of the transition $L$ and the polarization of the light $\sigma$. In this case $C_{ge}^2$ is obtained as \cite{Wense2020}
\begin{equation}
\begin{aligned}
C_{ge}^2=&(2F_e+1)(2F_g+1)(2I_e+1)\times\\
&\left\vert\biggl\lbrace\begin{matrix} I_g & L & I_e \\ F_e & J & F_g \end{matrix}\biggr\rbrace\right\vert^2 \left\vert \left(\begin{matrix} F_g & L & F_e \\ -m_{F_g} & \sigma & m_{F_e} \end{matrix}\right)\right\vert^2,\\
\end{aligned}
\end{equation}
where the curly bracket denotes the Wigner-6J and the curved bracket the Wigner-3J symbol, respectively. The definition of these symbols can be found, for example, in Ref.~\cite{Messiah}. In case that a stretched pair of nuclear hyperfine states is chosen, as proposed in the concept of Ref.~\cite{Campbell} for a single-ion nuclear clock, one obtains $C_{ge}^2=2/3$. Driving this sub-system will however require circular polarized light.\\[0.2cm] 
Assuming $C_{ge}=1$ for simplicity, the time required to reach saturation is about 20~s as shown in Fig.~\ref{excnucl_concept2}. The excitation probability amounts to 50\%, resulting in $5$ excited nuclei in case of resonance. In the search for the nuclear excitation the mode spacing of $77$ MHz has to be bridged. With a mode bandwidth of $490$ Hz, this leads to $1.57\cdot10^5$ scan steps. Assuming that each scan step takes a time of 20~s, the time required to bridge the mode spacing amounts to $3.14\cdot10^6$~s or 36.4 days. As all $1.2\cdot10^5$ comb modes are used in parallel, the energy range scanned in this way equals $9.2$~THz or $0.038$~eV. For this reason, it appears realistic to use the frequency comb in the search for the nuclear transition in Paul a trap, as soon as the energy has been constrained to an uncertainty which corresponds to the total bandwidth of the comb. In case of successful excitation, the isomeric energy could be constrained to a few 100~Hz of precision, providing the basis for the development of a single-ion nuclear clock. The main parameters of the concept are listed in Tab.~\ref{Table_concept2}.

\begin{table}[t]
\begin{center}
\caption{Main results for laser irradiation of $10$ trapped $^{229}$Th$^{3+}$ ions.}
\begin{tabular}{lc}
\hline\noalign{\smallskip}
Description & Value \\
\noalign{\smallskip}\hline\noalign{\smallskip}
Time per scan step & 20 s \\
Number of excited nuclei per scan step & 5  \\
Number of scan steps for 0.038 eV interval & $1.57\cdot10^5$  \\
Time required to scan 0.038 eV & 36.4 days  \\
\noalign{\smallskip}\hline
\end{tabular}
\label{Table_concept2}
\end{center}
\end{table}

\subsection{The potential for driving nuclear Rabi oscillations}
\label{concept3sec}
\noindent For the single-ion nuclear clock, it is the goal to irradiate an individual, laser-cooled $^{229}$Th$^{3+}$ ion with a single mode of a frequency comb with extraordinary narrow bandwidth \cite{Campbell}. Ideally, Rabi oscillations should be generated, in order to allow for the Ramsey interrogation scheme in the nuclear clock concept \cite{Ludlow}. This would lead to a quantum-projection-noise limited Allan deviation, which is a requirement for the highly stable operation of a single-ion nuclear clock. For the following it is assumed that the same frequency comb already discussed in the previous sections is used and focused to a diameter of 3~$\mu$m to irradiate a single ion, however, the bandwidth of an individual comb mode is narrowed down by a factor of $\sim500$ to 1~Hz. In this case it is shown that nuclear Rabi oscillations could be generated. The input values for Eq.~(\ref{Torrey}) are listed in Tab.~\ref{input_concept3}.

\begin{table}[t]
\begin{center}
\caption{Values of variables used for the calculation of nuclear Rabi oscillations.}
\begin{tabular}{ccl}
\hline\noalign{\smallskip}
Variable  & Value & Comment \\
\noalign{\smallskip}\hline\noalign{\smallskip}
$I$ & $0.14$ W/cm$^{2}$ & Single-mode focused to $\varnothing$ 3 $\mu$m\\
$\Gamma_L$ & $2\pi\cdot1$ Hz & Larger than nuclear linewidth\\
$\omega_0$ & $2\pi\cdot2.0$ PHz & Corresponding to 8.3 eV energy\\
$\Gamma_\gamma$ & $10^{-4}$ Hz & Estimated from theory \\
$\alpha_\text{eb}$ & $50$ &  EB decay for Th$^{3+}$ ions \cite{Mueller2017} \\ 
$N_0$ & $1$ & Individual laser-cooled  ion\\
\noalign{\smallskip}\hline
\end{tabular}
\label{input_concept3}
\end{center}
\end{table}

\begin{figure}[t]
\begin{center}
\includegraphics[totalheight=4.5cm]{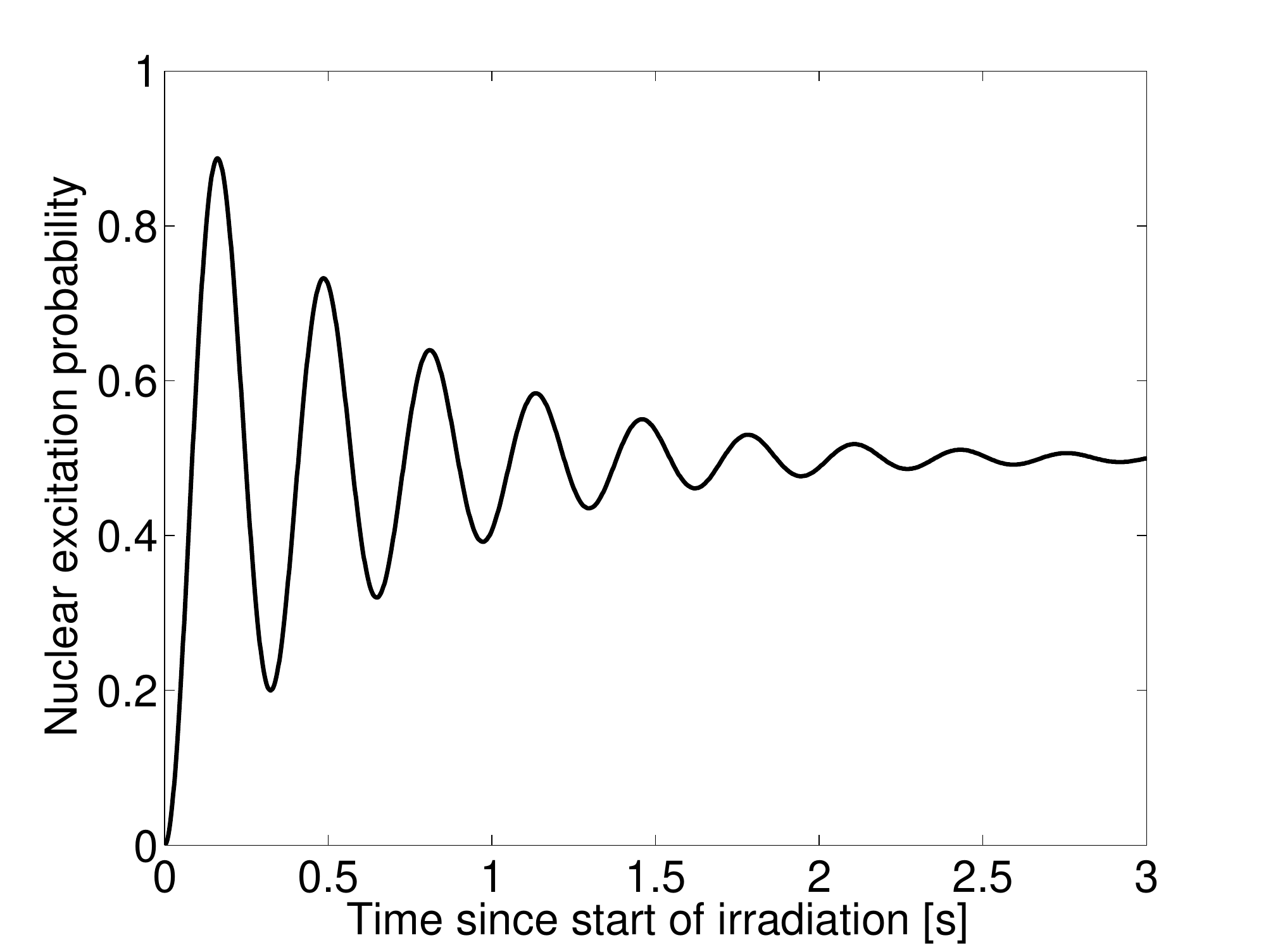}
\caption{\small Nuclear excitation probability as a function of time for a single $^{229}$Th$^{3+}$ ion in a Paul trap. Nuclear Rabi oscillations with a frequency of about 3~Hz become visible. Eq.~(\ref{Torrey}) was used with $C_{ge}=1$ and input values listed in Tab.~\ref{input_concept3}.}
\label{excnucl_concept3}
\end{center}
\end{figure}

\noindent The nuclear excitation probability as a function of time is shown in Fig.~\ref{excnucl_concept3}. Again, $C_{ge}=1$ was used and the result is not significantly affected by the large EB decay rate as long as $\Gamma\ll\Gamma_L=2\pi\cdot1$~Hz holds. This is the case for $\alpha_\text{eb}\ll 6.3\cdot10^4$, which will certainly be fulfilled. Based on Eq.~(\ref{Torrey}) the $\pi$-pulse duration is determined to $\approx170$~ms. The excitation probability after the $\pi$-pulse is about 90\%. Under the assumption that 1 second read-out time per scan step is required in order to observe the excitation via the double-resonance method, the time for scanning the $77$~MHz mode spacing is prohibitively long. For a bandwidth of $1$~Hz the number of scan steps amounts to $7.7\cdot10^7$. This corresponds to 2.4 years of scanning time in order to cover the energy range of 0.038 eV. However, in the case that the energy has been previously constrained by the concept described in Sec.~\ref{multipleions} to a few 100~Hz, a maximum number 500 scan steps would be required in the search for the resonance, leading to a net scanning time of 500~s. The main parameters of the concept are listed in Tab.~\ref{Table_concept3}.

\begin{table}[t]
\begin{center}
\caption{Main parameters for driving nuclear Rabi oscillations.}
\begin{tabular}{lc}
\hline\noalign{\smallskip}
Description & Value \\
\noalign{\smallskip}\hline\noalign{\smallskip}
$\pi$-pulse duration & $\sim$170 ms  \\
Excitation probability  & $\sim90\%$ \\
Time per scan step & 1 s \\
Number of scan steps for 500 Hz interval & $500$  \\
Time required to scan 500 Hz & 500 s  \\
\noalign{\smallskip}\hline
\end{tabular}
\label{Table_concept3}
\end{center}
\end{table}

\section{\label{sec3} Conclusion and outlook}
\noindent A new concept for narrow-band direct nuclear laser spectroscopy using the 7th harmonic of an Yb-doped fiber laser frequency comb is presented. The concept makes use of existing laser technology and would pave the way for narrow-band laser spectroscopy of $^{229\text{m}}$Th and the development of a solid-state nuclear optical clock.\\[0.2cm]
It is proposed to use the laser light for the irradiation of a 10~nm thin $^{229}$Th layer deposited as ThO$_2$ on a target substrate. In case of resonance, the successful excitation of the nuclear state can be probed via the detection of the internal conversion electrons emitted shortly after laser irradiation. The experiment appears to be advantageous compared to direct nuclear laser spectroscopy of $^{229}$Th ions in a Paul trap in terms of measurement time and absolute number of nuclear excitations.
The mode spacing of the frequency comb of 77~MHz could be scanned within 100~s, thereby effectively probing an energy range of 9.2~THz (0.038~eV), as all $1.2\cdot10^5$ comb modes are used in parallel in the search for the nuclear excitation. In case of success, the isomer's energy uncertainty would improve by a factor of $10^6$ down to about 100~MHz when compared with the free atom or ion. In addition, it would be straight-forward to develop an IC-based solid-state nuclear clock making use of this detection scheme, by stabilizing the comb mode to the nuclear transition. It is argued that this concept may be advantageous compared to the crystal-lattice nuclear clock approach, as it does not require the suppression of non-radiative decay channels, but makes instead use of these. An IC-based solid-state nuclear clock would have the same potential for the observation of time-variations of fundamental constants like a crystal-lattice nuclear clock \cite{Peik,Flambaum,Rellergert}.\\[0.2cm]
In addition, two experiments using the same frequency comb for excitation of laser-cooled $^{229\text{m}}$Th$^{3+}$ ions in a Paul trap were discussed for reasons of their importance: In the concept described in Sec.~\ref{multipleions}, multiple $^{229}$Th$^{3+}$ ions in a Paul trap are irradiated in parallel. This experiment is important, as it would allow constraining the isomeric energy value of individual $^{229}$Th$^{3+}$ ions by further six orders of magnitude, from 100 MHz to about 100~Hz, as required for the development of a single-ion nuclear clock. This cannot be achieved in a solid-state environment, as hyperfinestructure-shifts of the order of several 100~MHz are expected to occur. In case that the isomeric energy has been constrained by other methods to 9.2 THz (0.038~eV), corresponding to the total bandwidth of the frequency comb used for excitation, the required net scanning time would amount to 36.4 days. The concept can therefore be considered as realistic as soon as the isomer's energy value has been constrained by a further factor of 10 compared to the best current value.\\[0.2cm]
Finally, in Sec.~\ref{concept3sec} it is assumed that a single mode of the frequency comb is narrowed down to a bandwidth of 1~Hz. It is shown, that in this case it would be possible to drive nuclear Rabi oscillations with the laser system under consideration. Driving Rabi oscillations is important for the development of a single-ion nuclear clock, as they allow us to use the Ramsey interrogation scheme \cite{Ludlow}, thereby reducing the Allan deviation to the quantum-projection-noise (QPN) limit. This can be considered as a central requirement in order to take full advantage of the high expected accuracy of a single-ion nuclear clock. While conceptually possible, driving nuclear Rabi oscillations will require significant experimental effort and much improved constraints on the isomeric energy value as obtained by the concept discussed in Sec.~\ref{multipleions}. These experiments can be envisaged as follow-up investigations to the experiment discussed in Sec.~\ref{sec2}.\\[0.3cm]

\section{Acknowledgments}
\noindent We acknowledge discussions with B. Seiferle, G. Porat, S. Schoun, C. Heyl, D. Renisch, G. Kazakov, P. Bilous, A. Pálffy, F. Karpeshin, J. Weitenberg, O. Pronin, I. Pupeza, T. Udem, E. Peik, S. Stellmer, T. Schumm, C.E. Düllmann, T. Schibli, P.G. Thirolf and J. Ye. C. Zhang acknowledges the Graduate Research Assinstantship from J. Ye's group in JILA. Part of this work was submitted as a research proposal by L. v.d.Wense for a Humboldt fellowship, which has been granted. L. v.d.Wense is grateful to the Humboldt Foundation for this support. Further, he would like to thank J. Ye for accepting him as a Humboldt fellow and P.G. Thirolf for eight years of successful joint work. This work was funded by the European Union's Horizon 2020 research and innovation programme under grant agreement 664732 ``nuClock" and by DFG grant No. Th956/3-2 as well as by the LMU Mentoring program.

\end{document}